\documentclass[11pt,a4paper]{article}
\pdfoutput=1
\usepackage{jheppub}

\usepackage{color}
\usepackage{amsmath}
\usepackage{pifont}

\usepackage{verbatim}   
\usepackage{subfigure}       
\usepackage{acronym}  
    
\usepackage{amsfonts}
\usepackage{amssymb}
\usepackage{mathrsfs} 
\usepackage{graphicx}
\usepackage{multirow} 
 \usepackage{slashed}

\newcommand{\uo}{${{\rm U \left(1 \right)}}$}
\newcommand{\suw}{${{\rm SU \left(2 \right)}}$}
\newcommand{\suc}{${{\rm SU \left(3\rm \right)}}$}

\newcommand{\GeV}{{\, {\rm GeV}}}
\newcommand{\TeV}{{\, {\rm TeV}}}

\def\beq{\begin{equation}}
\def\eeq{\end{equation}}
\def\bea{\begin{eqnarray}}
\def\eea{\end{eqnarray}}
\def\bitem{\begin{itemize}}
\def\eitem{\end{itemize}}
\newcommand{\bec}{\begin{center}}
\newcommand{\eec}{\end{center}}
\newcommand{\ba}{\begin{array}}
\newcommand{\ea}{\end{array}}

\def\bar#1{\overline{#1}}

\def\bra#1{\left\langle #1\right|}
\def\ket#1{\left| #1\right\rangle}

\def\inv{^{\raise.15ex\hbox{${\scriptscriptstyle -}$}\kern-.05em 1}}
 
\def\lbar{{\lower.35ex\hbox{$\mathchar'26$}\mkern-10mu\lambda}}

\def\to{\rightarrow}

\let\<=\langle
\let\>=\rangle

\let\+=\uparrow

\begin{document}

\title{Dirac vs Majorana gauginos at a 100 TeV collider} 
\author[a]{Giovanni Grilli di Cortona,}
\author[b]{Edward Hardy,}
\author[c]{and Andrew J. Powell}
\emailAdd{ggrilli@sissa.it,ehardy@ictp.it,andrew.powell2@physics.ox.ac.uk}
\affiliation[a]{SISSA - International School for Advanced Studies, and INFN - Sezione di Trieste, Via Bonomea 265, 34136, Trieste, Italy}
\affiliation[b]{Abdus Salam International Centre for Theoretical Physics,
Strada Costiera 11, 34151, Trieste, Italy}
\affiliation[c]{Rudolf Peierls Centre for Theoretical Physics, 1 Keble Road, Oxford, OX1 3NP, UK}

\abstract{
We compare the prospects for observing theories with Majorana or Dirac gauginos at a future 100 TeV proton--proton collider. Calculating the expected discovery and exclusion regions, we find that for heavy gluino masses the squark discovery reach is significantly reduced in Dirac gluino models relative to the Majorana case. However, if the squark and gluino masses are close the reach is similar in both scenarios. We also consider the electroweak fine tuning of theories observable at such a collider, and the impact of existing constraints from flavour and CP violating observables. Models with Majorana gluinos that are fine tuned to less than one part in $10,000$ can typically be discovered or excluded, and Dirac models with tuning of one part in $1,000$ can be probed. The flavour structure of Majorana models is highly constrained if they have observable squarks, while $\mathcal{O}(1)$ violation is possible in accessible Dirac models. In both cases new sources of CP violation must be very suppressed. Future collider searches 
can also give important information on possible dark matter candidates. We study the relation of this to indirect and direct detection searches, and find that if dark matter is a neutralino, a 100 TeV collider could probe the viable dark matter candidates in large classes of both Dirac and Majorana models.

}

\maketitle

\section{Introduction}
\label{sec:intro}

Despite negative results from searches at the LHC, low scale supersymmetry (SUSY) remains a well motivated scenario for physics beyond the Standard Model (SM). Theories consistent with observations require a moderate fine tuning, but if this is accepted they can explain the remainder of the hierarchy between the electroweak (EW) and the Planck scale, and can also lead to viable dark matter candidates and gauge unification. Consequently there has been significant interest in the prospects for discovering supersymmetry at a future hadron collider with center of mass energy in the region of 100 TeV. Studies have examined the discovery reach for squarks and gluinos \cite{Stolarski:2013msa,Cohen:2013xda,Cohen:2013zla,Fowlie:2014awa,Cohen:2014hxa,Jung:2013zya}, as well as for neutralinos and the interplay of such searches with dark matter direct detection \cite{Zhou:2013raa,Low:2014cba,Cirelli:2014dsa,Gori:2014oua,Acharya:2014pua,Bramante:2014tba,diCortona:2014yua}.

In light of challenges faced by the Minimal Supersymmetric Standard Model (MSSM), it is also interesting to consider non-minimal implementations of SUSY. In this paper we look at models with Dirac gaugino masses, and compare these with the standard Majorana case. Models where gauginos have Dirac masses were considered early in the study of supersymmetric theories  \cite{Fayet:1978qc,Polchinski:1982an,Hall:1990hq}, and have received renewed interest because of potential phenomenologically appealing features \cite{Fox:2002bu,Kribs:2007ac,Kribs:2012gx}. These include possibly weakening collider limits and flavour constraints, and reducing EW fine tuning, compared to Majorana models.  In particular, our study consists of two parts. First, we revisit the discovery reach of a future 100 TeV collider for strongly coupled states in theories with Majorana gluinos, and extend this to cases with Dirac gluinos. Secondly, we consider the consequences of the projected collider reach for model building in the Majorana 
and Dirac scenarios, asking what  flavour, dark matter, and fine tuning features a model must have in order that it is both allowed by current observations and discoverable at a 100 TeV collider.

Our simulation of the collider reach is carried out in Section \ref{sec:search}. Focusing on a simplified model with gluinos, the first and second generation squarks, and a light neutralino, we compare the SUSY production cross sections for scenarios with Majorana and Dirac gluino masses. We then calculate the expected discovery and exclusion regions in these models for a 100 TeV proton--proton collider  with $3\text{ ab}^{-1}$ integrated luminosity, and comment on the impact of this being increased to $30\text{ ab}^{-1}$. In parts of parameter space where the gluino mass is heavy compared to the squarks, squark-squark production is dominant. This is suppressed in Dirac compared to Majorana models due to the lack of a chirality flip, leading to large differences in the expected reach. However, if the squark and gluino masses are comparable, or the squarks are heavier the reach is similar in both cases.
We also briefly discuss models with the stops light compared to the first two generation sfermions, and the possibility of observing sgluons.

In Section \ref{sec:model} we consider the collider reach in light of the relations between soft masses that are well motivated by UV completions of models with Dirac gauginos, although we do not commit ourself to a particular model. The mass of the lightest physical Higgs is typically not a free parameter in supersymmetric theories, instead being fixed by the values of the soft masses. Therefore we also study how the observation of a Higgs at $125\,\GeV$ with close to SM couplings affects model building. Additionally, we estimate the degree of EW fine tuning in models that will be probed by future colliders.

Supersymmetric models often give extra contributions to rare SM flavour and CP violating processes. The rate of these is observed to be close to the SM predictions, and as a result theories with superpartners accessible to the colliders must have soft term spectra such that the dangerous processes are sufficiently suppressed. In some mediation mechanisms, such as gauge mediation, this can be easily accommodated. However in others, especially string theory completions of gravity mediation, it remains a pressing issue. In Section \ref{sec:fl}, we revisit these bounds and study how constrained the flavour and CP sectors of a model must be if it has superpartner masses directly observed at a 100 TeV collider.

Another feature of supersymmetric theories is that a dark matter candidate is present in many motivated models, and if the mediation scale is high this is typically the lightest neutralino. In Section \ref{sec:dm} we consider the dark matter candidates in Majorana and Dirac models, and the relation between searches at a 100 TeV collider and direct and indirect detection. In theories with an R-symmetry, direct detection experiments already rule out many neutralino dark matter candidates, and a 100 TeV collider can efficiently probe the remaining possibilities. In the case of wino-bino dark matter this is through searches for charged winos. Meanwhile searches for gluinos can strongly constrain the gluino bino mass ratio in models in which a bino coannihilates with sleptons. If the neutralino sector does not have an R-symmetry there are more dark matter candidates, but a 100 TeV collider is sensitive to a significant proportion of viable models.
Finally, in Section \ref{sec:con} we conclude.

\section{Strongly coupled states at 100 TeV collider} \label{sec:search}

Hadron colliders are very efficient at producing strongly interacting states, and even though coloured superpartners are often amongst the heaviest they are consequently important for the discovery or exclusion of theories. We therefore study the production cross sections of these at a 100 TeV proton--proton collider. Further, by employing a squark-gluino-neutralino simplified model, and scanning over squark and gluino masses, we obtain the expected discovery and exclusion reach of such a collider. The Dirac or Majorana nature of gluino masses leads to significant differences in the production rates in some parts of parameter space \cite{Choi:2008pi,Heikinheimo:2011fk,Kribs:2012gx,Kribs:2013eua} and we highlight the effects of these. LHC searches already set stringent limits on SUSY models \cite{TheATLAScollaboration:2013fha,CMS:2013gea} and these can be recast to give significant constraints on models of Dirac gluinos. The bounds on the first two generation squark masses from 8 TeV data are found to be roughly in the 
region of $800\,\GeV$ for Dirac gluino masses of around $5~\TeV$, and are expected to reach somewhere in the region of $1.2\,\TeV$ with 14 TeV data \cite{Kribs:2012gx,Kribs:2013eua}.

\subsection{Production cross sections}\label{sec:prod}

Motivated by UV theories of Dirac gluinos, to be discussed in Section \ref{sec:model}, we consider two benchmark patterns of soft terms, one with $m_{\tilde{q}}=m_{\tilde{g}}$ and the other with $m_{\tilde{g}}=5m_{\tilde{q}}$ where $m_{\tilde{g}}$ is the gluino mass. Due to the parton content of the proton, the first two generation sfermions are produced far more readily than stops, provided the masses are not very hierarchical. Therefore we take $m_{\tilde{q}}$ to be a degenerate mass for these states, and neglect the production of stops. Our calculations are performed with MadGraph5 \cite{Alwall:2014hca}, and we have not included next to leading order (NLO) K-factors in this section. These are not yet known for Dirac gluinos and we are primarily concerned with the relative sizes of cross sections here. 
Cross sections for Majorana models at 100 TeV have previously been studied  (for example in \cite{Borschensky:2014cia}), while cross sections for Dirac models at an energy of 33 TeV are given in \cite{Kribs:2013oda}, and we find good agreement with these.

\begin{figure}[t]
\centering
\includegraphics[width=0.47\textwidth]{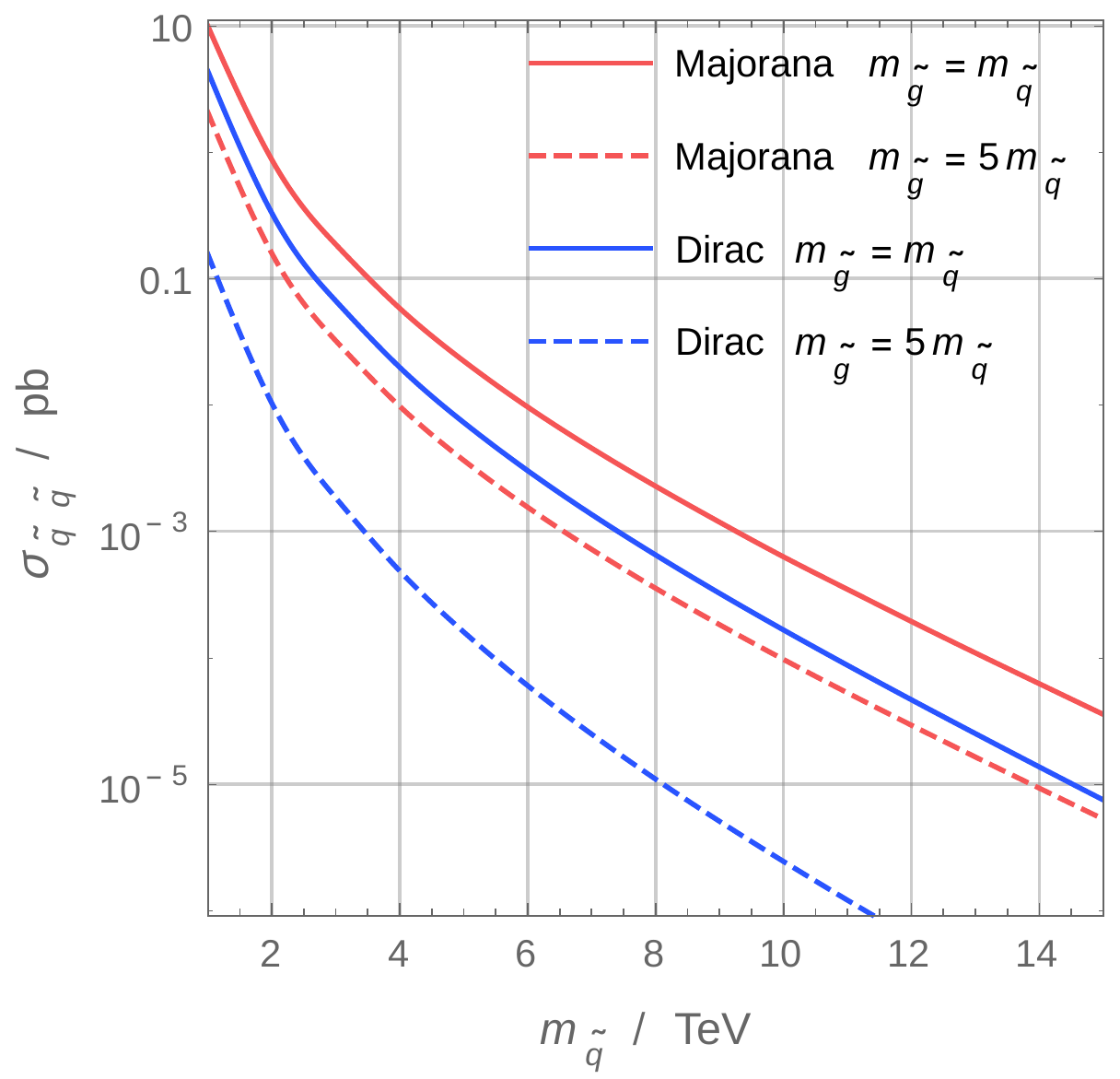}
\qquad
\includegraphics[width=0.47\textwidth]{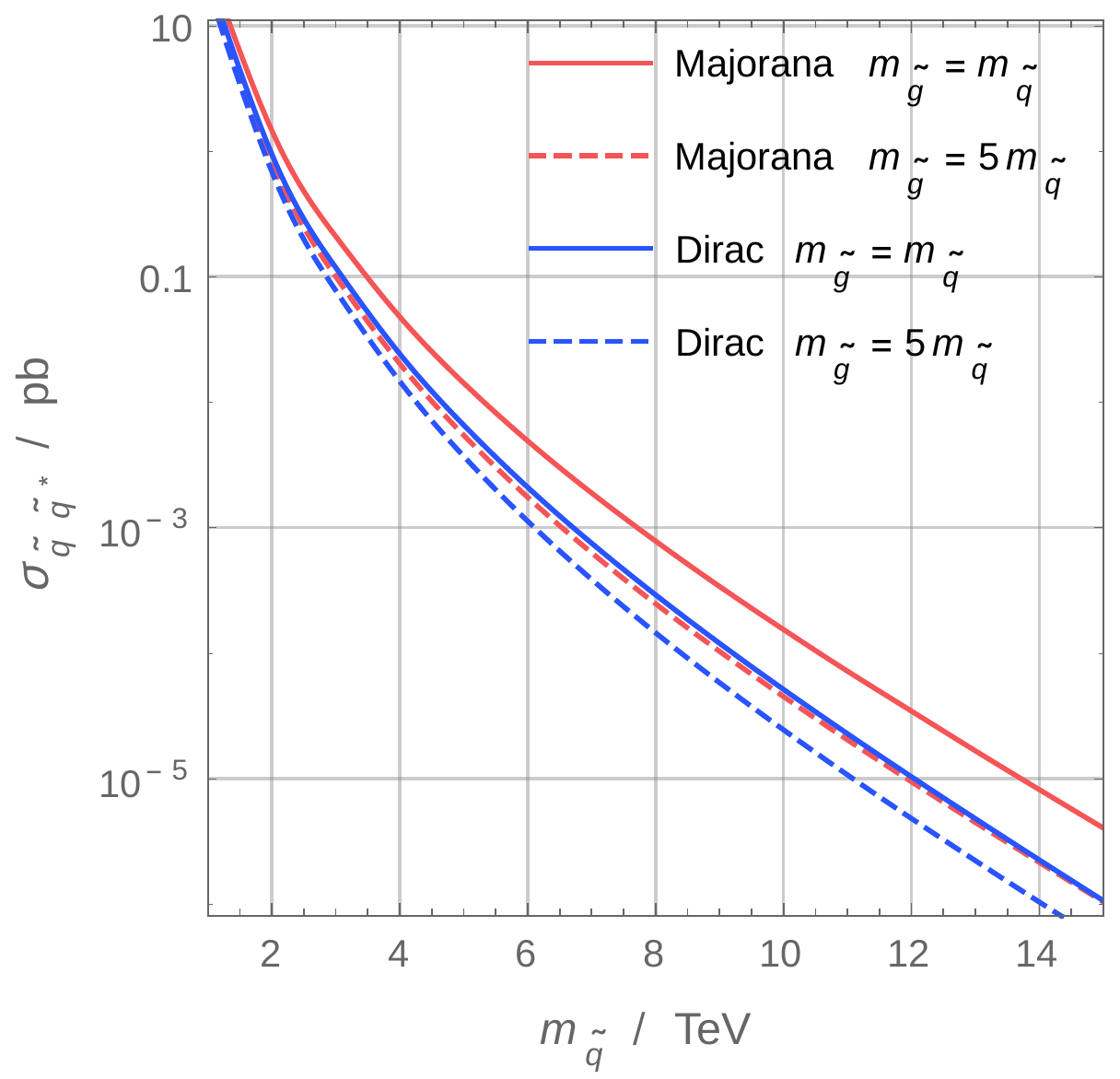}
\caption{{\bf \emph{Left:}} The squark-squark  production cross sections for Dirac (plotted in blue) and Majorana (red) gluinos as a function of the squark mass. We show the production cross sections for two benchmark relations between the soft parameters, $m_{\tilde{g}}=m_{\tilde{q}}$ (solid lines) and $m_{\tilde{g}}=5m_{\tilde{q}}$ (dashed lines).  {\bf \emph{Right:}} The squark-anti-squark production cross sections for the same models.}
\label{fig:sqsqcprod}
\end{figure}

In Fig.~\ref{fig:sqsqcprod} (left) we show the production cross sections for squark pairs for Majorana and Dirac gluinos. This production channel shows the most striking difference between the two cases. For Dirac gluinos only production of $(\tilde{q}_L\tilde{q}_R)$ is possible, with $(\tilde{q}_L\tilde{q}_L)$ forbidden due to the lack of an allowed chirality flip. In contrast, a chirality flip is possible with a Majorana gluino mass, and consequently this cross section is dramatically larger than in Dirac models \cite{Kribs:2013oda}.\footnote{The difference in the relative production of same chirality and opposite chirality squarks could even be used to distinguish Dirac from Majorana gluinos \cite{Heikinheimo:2011fk}.} 
In Fig.~\ref{fig:sqsqcprod} (right) the production cross sections for squark-anti-squark pairs is plotted. The cross section is only slightly reduced in  Dirac models because the dominant production mode is through an s-channel gluon. This production mode is independent of the gluino mass, reflected in the relatively small drop  in  cross sections between the models with gluino mass equal to the squarks and the models with a heavy gluino.

\begin{figure}
\centering
\includegraphics[width=0.47\textwidth]{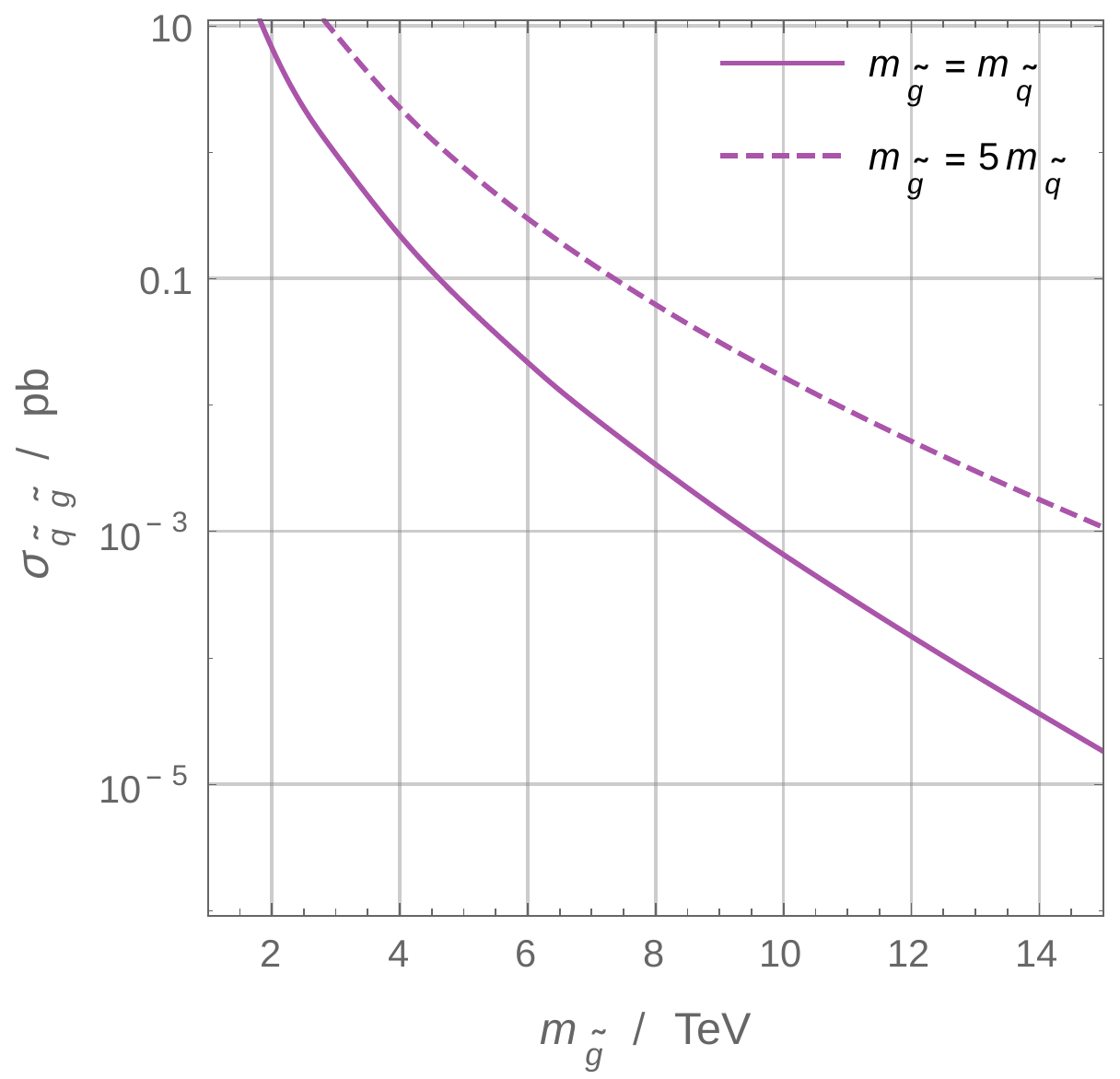}
\qquad
\includegraphics[width=0.47\textwidth]{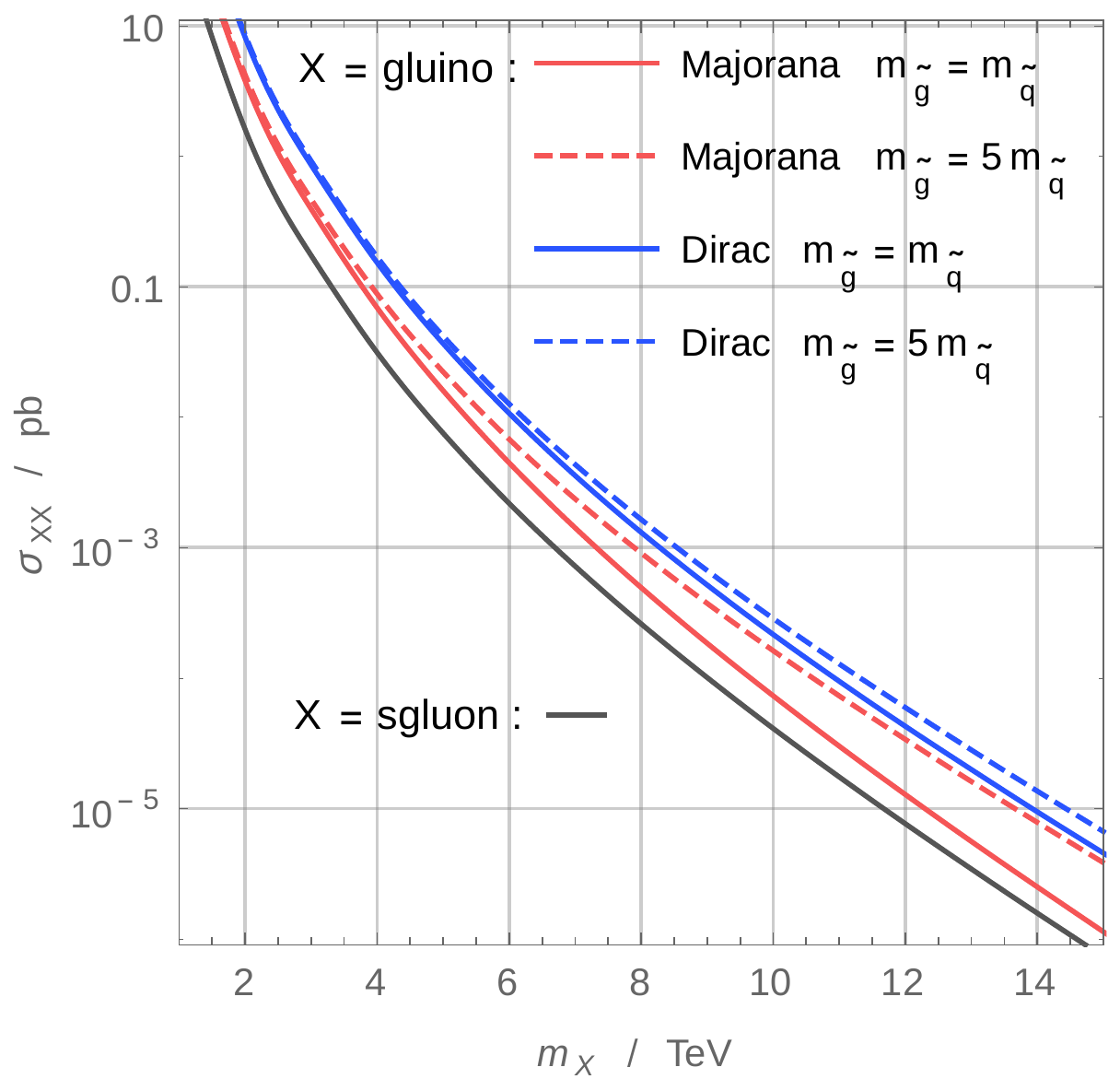}
\caption{{\bf \emph{Left:}} The gluino-squark production cross section for the two soft mass relations, as a function of the gluino mass. To a good approximation this is identical for the Dirac and Majorana scenarios. For the $m_{\tilde{g}}=5m_{\tilde{q}}$ curve, the plotted region corresponds to light squarks, and gluino-squark production is entirely negligible for models with such a mass relation if the squarks are heavier than a few TeV. {\bf \emph{Right:}} The gluino pair production cross sections, for the Dirac (blue) and Majorana (red) benchmark models as a function of the gluino mass. We also show the production cross section for pair production of sgluons (plotted in black), as a function of the sgluon mass, which is approximately independent of the other soft parameters in the model.}
\label{fig:glglprod}
\end{figure}

In Fig.~\ref{fig:glglprod} (left) we show the gluino-squark production cross section. To a good approximation this is the same for both Dirac and Majorana models. If the squarks are relatively heavy compared to the gluino, gluino pair production is the dominant source of superparticles. The cross section for this is plotted in Fig.~\ref{fig:glglprod} (right), and here the only difference between the models is the additional degrees of freedom in the Dirac gluino model. This simply enhances the production cross section of gluino pairs by a factor of a few.

Therefore the main differences in exclusion and discovery reach will be in regions of parameter space where gluino pair production, or squark pair production dominates. With relatively heavy squarks, gluino pair production will be most important and the Dirac model will lead to higher potential exclusion.

Comparing the cross sections in Figs.~\ref{fig:sqsqcprod} and \ref{fig:glglprod}, we see that when the squark masses are equal to the gluino mass squark-gluino production will be large, and in Majorana models there will be also be a comparable squark-squark production cross section, while this is suppressed in Dirac models. Consequently, the collider reach is expected to be roughly similar in the two cases. For larger gluino masses, the 
squark-gluino cross section decreases fast, and squark-squark production dominates. In this part of parameter space the exclusion and discovery potential will be drastically reduced in the Dirac model compared to the Majorana model. Eventually, in the decoupling limit where the gluino is so heavy that it is completely removed from the spectrum, squark-anti-squark production is dominant and the two models have the same production cross section.

As well as squarks and gluinos, Dirac gluino models feature another new coloured state. This is the sgluon $\phi$,  a complex scalar in the adjoint of \suc. The sgluon mass depends on the UV completion of the theory, but is often comparable in mass to the gluino.
Over most of parameter space sgluons are dominantly pair produced with single production significantly suppressed \cite{Plehn:2008ae}, and to a good approximation the cross section for this is independent of the other soft parameters of the theory. In Fig.~\ref{fig:glglprod} (right), we therefore show the leading order production cross section for pair production, assuming the scalar and pseudo-scalar components are degenerate in mass (NLO corrections for sgluon production have been obtained in \cite{GoncalvesNetto:2012nt}, and including these does not qualitatively affect our conclusions). It can be seen that sgluons are pair produced less frequently than gluinos unless they are substantially lighter.

\subsection{Discovery and exclusion reach}

To determine the expected discovery and exclusion reach of a 100 TeV proton--proton collider, we study a simplified  squark-gluino-neutralino model. We set the neutralino mass to $100~\GeV$, and scan over the masses of the degenerate first and second generation squarks, and the gluinos, with all other superpartners decoupled. The signal of such a model, jets and missing energy, is a classic search for supersymmetry, and the main SM backgrounds to this are W/Z boson + jets, and $t\bar{t}$ + jet production.

Our simulation is performed using the SARAH Dirac Gauginos model \cite{Staub:2012pb}. 
We produce parton level events for the production channels $(\tilde{g}\tilde{g})$, $(\tilde{q}\tilde{g})$, $(\tilde{q}\tilde{q})$, and $(\tilde{q}\tilde{q}^*)$ in MadGraph5 \cite{Alwall:2014hca}. Decay, showering and hadronisation is carried out with Pythia6.4 \cite{Sjostrand:2006za}, and we use the Snowmass detector card \cite{Anderson:2013kxz} in Delphes \cite{deFavereau:2013fsa} to perform our detector simulation. Given how rapidly the cross sections fall with increasing superpartner masses the order one uncertainties on the properties of a future detector are expected to lead to relatively small changes to our results. To obtain the SM backgrounds we use the publicly available Snowmass results \cite{Avetisyan:2013onh}. We assume that there are 20\% systematic uncertainties associated with the SM backgrounds (this may be a fairly cautious estimate and we comment on the effect of altering it). Further, we 
compute the expected number of events at a 100 TeV collider assuming, somewhat conservatively, $3~{\rm ab}^{-1}$ worth of luminosity. Total integrated luminosities of up to $30~{\rm ab}^{-1}$ have been suggested as a suitable aim for a 100 TeV collider, and if this is achieved a significant increase in mass reach is possible \cite{Hinchliffe:2015qma,Arkani-Hamed:2015vfh}, and we discuss this later.

To determine the discovery reach we first perform a series of preselection cuts to remove the majority of the SM background. For this we follow the choices of the Snowmass study \cite{Cohen:2013xda}, although further optimisation may be possible and could lead to a small improvement in reach. We require all jets have at least $30~\GeV$ transverse momentum, otherwise we ignore that jet. At least four jets with transverse momentum greater than $60\,\GeV$ are also required, and we perform a cut on the missing energy $E_{T,\,\text{miss}}$ and $H_t$, the scalar sum of the final state jets, of $E_{T,\, \text{miss}}^2/H_T> 225 ~\GeV$. Further, we also demand there are no leptons in the final state. We then scan over square cuts on $E_{T,\,\text{miss}}$ and $H_T$, with the final cuts chosen so as to maximise the signal significance, i.e. maximising, see \cite{Ellis:2015xba}
\begin{equation}
\sigma=\frac{S}{\sqrt{1+B+\gamma^2 B^2 + \delta^2 S^2}}~,
\end{equation}
where $S$ and $B$ are the number of signal and background events, and $\gamma$ and $\delta$ are the assumed systematic uncertainties on the background and signal events. 

NLO K-factors $K= \sigma_{{\rm NLO}} / \sigma_{{\rm LO}}$ are calculated for Majorana gluinos using Prospino2.1 \cite{Beenakker:1996ch}. These factors have not been calculated for the Dirac case. As an estimate of the effect of going to NLO we apply the K-factors calculated with a Majorana gluino to both models. We do not generate events with any additional parton level jets, nor do we include the effects of pile up. Pile up could be a substantial challenge at a 100 TeV collider \cite{Barletta:2013ooa,Skands:2013asa}, but it has been argued that the present analysis is not likely to be very sensitive to its effects \cite{Cohen:2013xda}.

We study the discovery and exclusion potential in both Majorana and Dirac gluino models, and scan over gluino masses between $8$   and $24~\TeV$, and squark masses between $6$ and $20~\TeV$. For squarks much heavier than the gluino, the main decay mode for the gluino is to a $q\,\bar{q}\,\chi^0$ final state, whilst the squark decays to $\tilde{g}\,\chi^0$. For the opposite case, the squark decays to a $q\,\chi^0$ final state, while the gluino decays to $\tilde{q}\,\bar{q}$.

\begin{figure}
\begin{center}
 \includegraphics[width=0.47\textwidth]{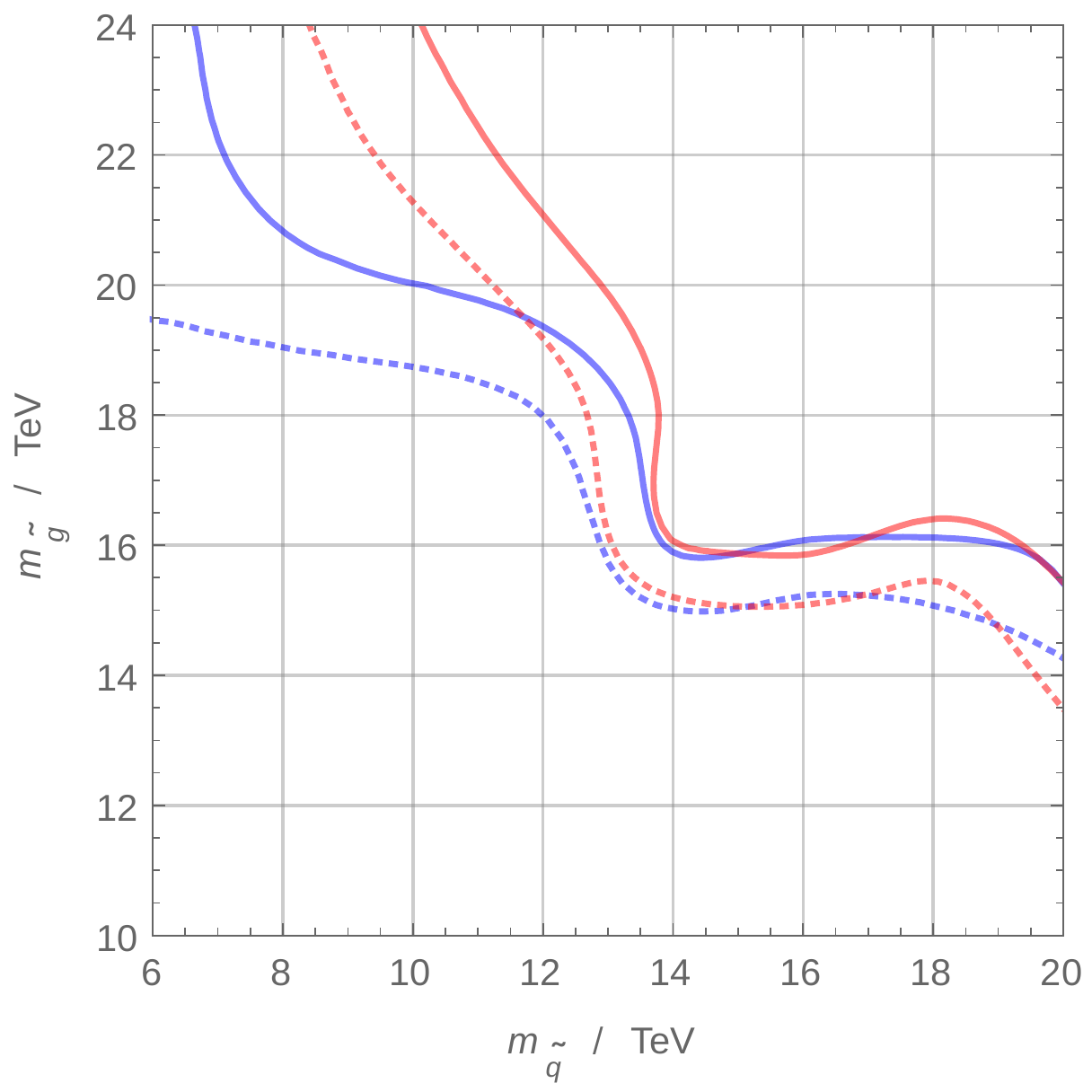}
 \qquad
\includegraphics[width=0.47\textwidth]{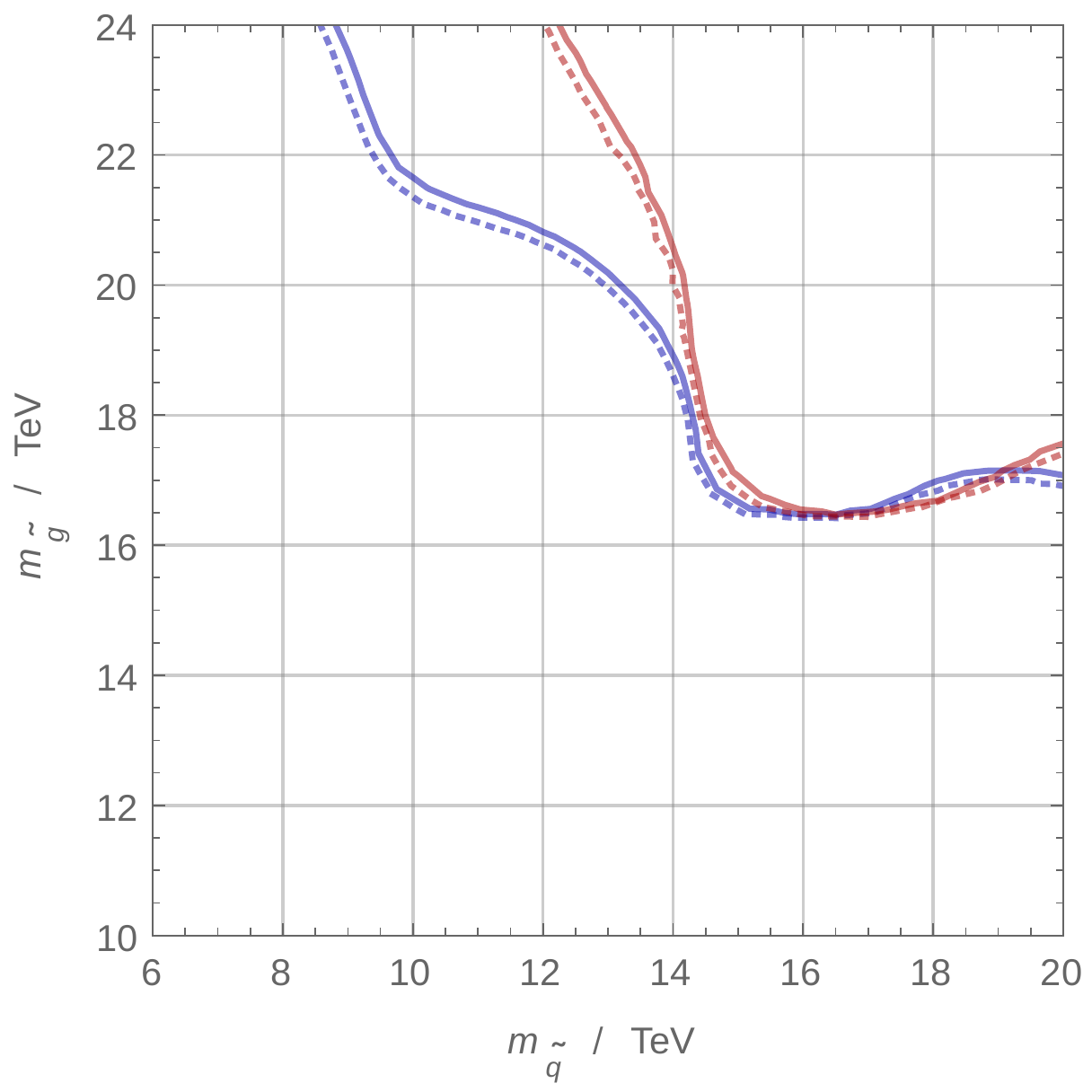}
\caption{{\bf \emph{Left:}} The 5$\sigma$ discovery potential of the Dirac (blue) and Majorana (red) gluino-squark-neutralino model including the Majorana K factors for both models at a 100 TeV collider, for 5\%, and 15\% systematic uncertainty on the signal shown solid and dotted respectively. {\bf \emph{Right:}} The expected 95\% exclusion bounds for the same models.}
\label{fig:majoranadiscovery}
\end{center}
\end{figure}

\begin{figure}
\begin{center}
 \includegraphics[width=0.47\textwidth]{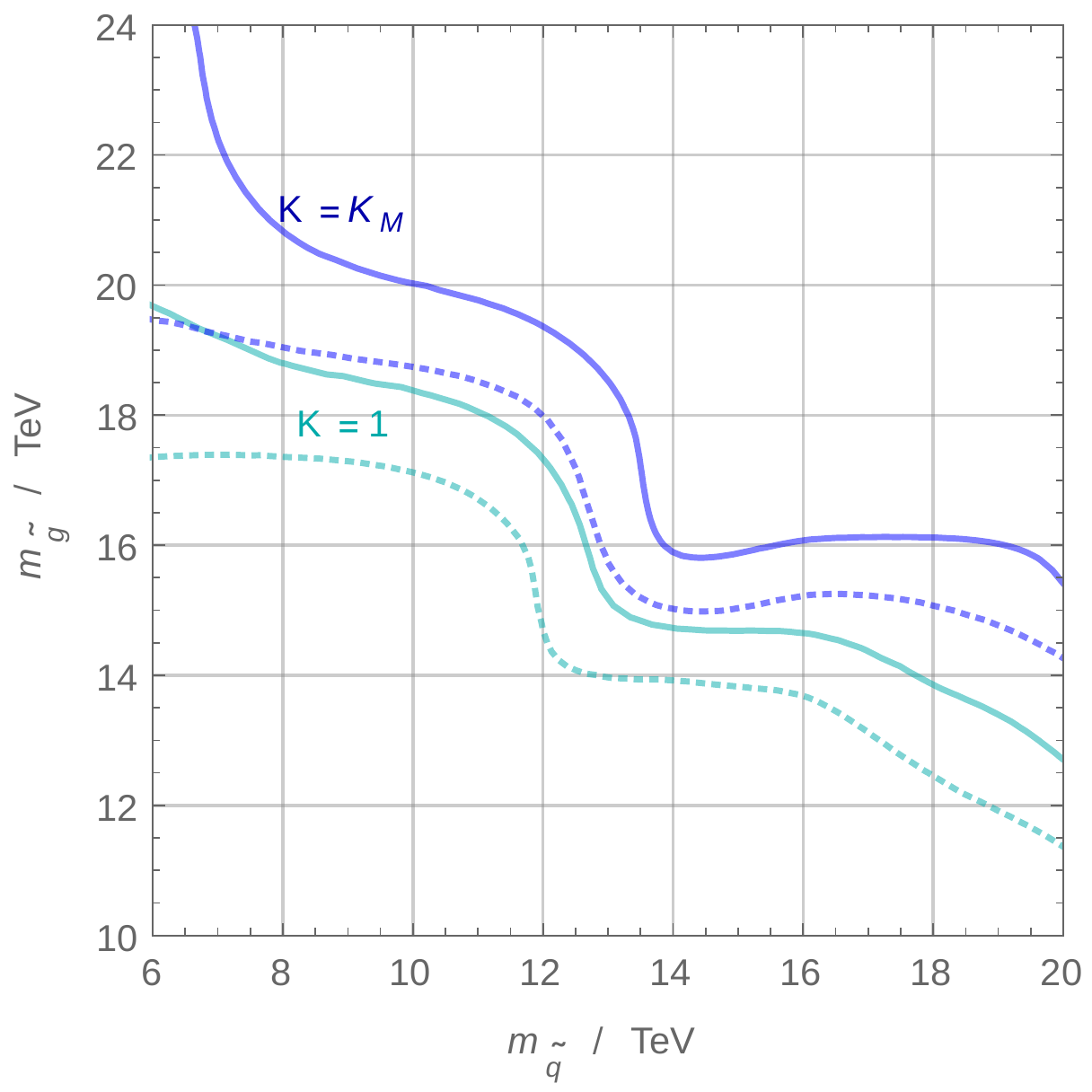}
 \qquad
\includegraphics[width=0.47\textwidth]{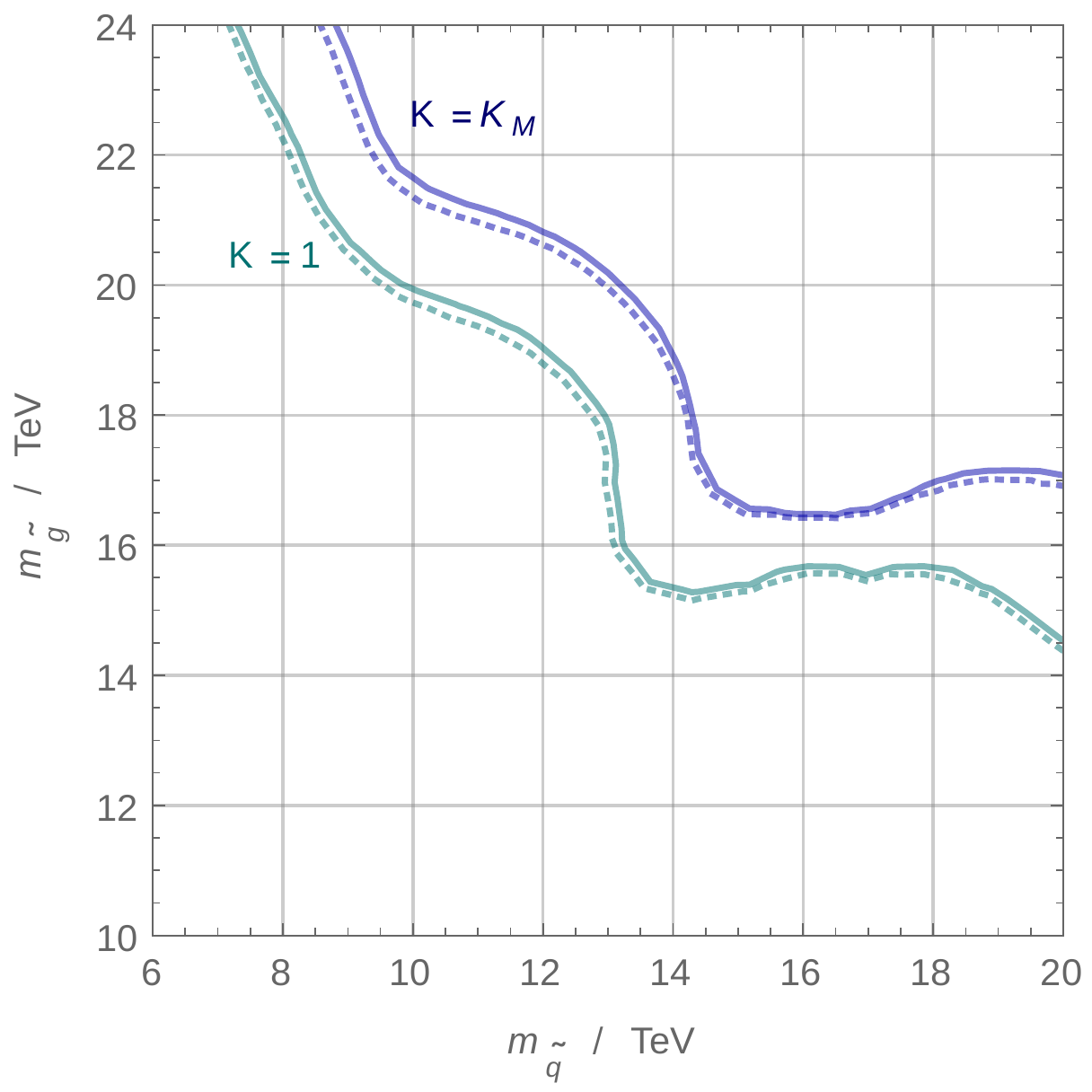}
\caption{{\bf \emph{Left:}} The 5$\sigma$ discovery potential of the Dirac gluino-squark-neutralino model including the Majorana k factors, dark blue, and with no K factor, light blue, for 5\%, and 15\% systematic uncertainty on the signal shown solid and dotted respectively. {\bf \emph{Right:}} The expected 95\%  exclusion for the same models.}
\label{fig:Keffect}
\end{center}
\end{figure}

Of primary interest to us is the difference between the Dirac and Majorana models, and this is plotted in  Fig.~\ref{fig:majoranadiscovery}. In the region of parameter space where the masses of the gluinos and squarks are comparable, we find both models have the same discovery potential. Similarly, if the squark mass is above the gluino mass then the discovery reach is comparable between the models. For larger still squark masses, much above the gluino mass, gluino pair production will dominate and in this case a Dirac model will have slightly higher exclusion reach. In both models a 100 TeV collider would be able to discover gluinos with masses up to approximately $15~\TeV$, and exclude gluino masses of approximately $17~\TeV$. 

The main difference comes in the region with heavy gluinos, with masses in the range $18$ to $24~\TeV$. In a Majorana model a 100 TeV collider can probe squark masses up to approximately $11~\TeV$, however a Dirac model only has sensitivity to masses up to $6~\TeV$. This is expected given the relative sizes of the production cross sections. In this mass region the dominant Majorana production mode is squark-squark pairs, but in a Dirac model this production mode is very suppressed. In theories of Dirac gluinos, the dominant production is instead through squark-anti-squark pairs, which has a smaller cross section leading to reduced sensitivity.

It can be seen from the figures that the systematic uncertainty on the signal leads to reasonable differences in the discovery potential, of order a few TeV, and the range of values shown is plausible given the uncertainties in typical LHC analyses. The effect of this uncertainty on the exclusion potential is relatively minor. Changing the systematic uncertainty on the SM background has little effect on the expected discovery and exclusion regions, with statistical uncertainty being the dominant source of error on the SM background.

General arguments, based on parton distribution functions, suggest that an order of magnitude increase in integrated luminosity from $3~\text{ ab}^{-1}$ to $30~\text{ ab}^{-1}$ could boost the reach for heavy particles by several $\TeV$ \cite{Hinchliffe:2015qma,Arkani-Hamed:2015vfh}. Repeating our analysis with a luminosity of $30~\text{ ab}^{-1}$, we find that this is indeed the case. In the heavy gluino part of parameter space, squark masses about $4~\TeV$ larger than before are probed, so that in the Dirac model squarks with masses between $8$ and $10~\TeV$ could be discovered depending on the signal uncertainty. If the squark and gluino masses are similar the reach is increased by about $2~\TeV$ in both masses.

Let us also note the impact of using NLO K-factors, which is shown in Fig.~\ref{fig:Keffect}. We have computed the K-factors using Prospino2.1 in a model with Majorana gluinos. NLO K-factors have not been calculated for Dirac gluino models to our knowledge. Thus the inclusion of these K-factors in the Dirac model should be thought of purely as an estimate of the impact of going to NLO, rather than a firm prediction. We see that the inclusion of K-factors has a dramatic effect on the discovery and exclusion potential of both models, and thus calculation of K-factors for Dirac gluino models is an important task if accurate predictions are to be made.

A shortcoming of our present work is that we have not included additional parton level jets, which is computationally expensive, in our simulation. The effect of this can be estimated by comparing our Majorana results with the Snowmass study \cite{Cohen:2013xda}, which includes up to two additional jets. Our results match the Snowmass study well in most of the parameter space, while there is some difference in the high gluino mass region. In particular, the Snowmass study finds sensitivity to $14~\TeV$ squarks for gluinos at masses around $20~\TeV$, falling to $12\TeV$ squarks for gluino masses of $24~\TeV$. In comparison we find a maximum discovery reach of $10$ to $12~\TeV$ squark masses for these gluino masses. It is likely that in this parameter region the lack of jets results in a signal that is not well separated from the background. However even for such spectra the numerical error introduced is relatively minor compared to, for example, uncertainties from the unknown luminosity a future collider 
might achieve.

As well as models with close to universal squark masses, it is well motivated to consider theories with stops relatively light compared to the first two generation squarks. In the case where the gluino is decoupled, the discovery potential of this model is no different to that studied in \cite{Cohen:2014hxa}. In a simplified model with a neutralino, it is found that pair production allows $5.5~\TeV$ stops to be discovered, while $8~\TeV$ stops can be excluded.

Differences between Dirac and Majorana models can arise if the gluino mass is not completely decoupled from the spectrum. In Dirac models the stop-anti-stop production cross section is only slightly suppressed compared to a Majorana model due to the main production mode being through an s-channel gluon. Thus if the gluino is out of kinematic reach and the only way to produce stops is through pair production, then we expect only a slight reduction in sensitivity in Dirac models compared to a Majorana model. However, this could change if  stops and gluinos are both kinematically accessible. Since gluino production is larger in Dirac models, stop production through gluino decay would be more important  in this case. As the stop masses are varied, it is expected that the bound on gluino masses will be at least as strong as that obtained when all squarks are decoupled, giving us a lower limit on the discoverable parameter space.  A dedicated analysis of 
these scenarios at a 100 TeV collider, along the lines of the study of LHC search carried out in \cite{Papucci:2011wy}, would be worthwhile.

 The discovery potential for sgluons depends on details of the pattern of superpartner masses. Decays to a squark-anti-squark pair are typically dominant if they are kinematically allowed, and otherwise decays to a pair of gluinos or a quark-antiquark pair though a loop of squarks can be significant. In the case of decays to quark-antiquark pairs, the rate is suppressed by the mass of the quark, so a substantial proportion of events involve top quarks. This leads to the interesting possibility of searching for events with same sign top quarks \cite{Plehn:2008ae}, which is reasonably easy to distinguish from SM backgrounds. Studies relevant to the LHC have been carried out \cite{Gresham:2007ri,Choi:2008ub,Martynov:2008wf,Choi:2009ue,Choi:2009jc,Fuks:2012im,Calvet:2012rk}, and assuming a model where sgluons decay mostly to tops, LHC searches rule out sgluon masses up to the region of $700\,\GeV$. Since, in most of the parameter space of motivated models, sgluons do not lead to the dominant discovery and 
exclusion potential we do not consider them further in our present collider simulations. However, if gluinos or squarks were to be discovered they would be a very exciting state to search for, not least because their presence would be a very strong hint that gluino masses were at least partially Dirac.

We stress that our analysis and simplified model nowhere near covers the full range of possible models and signatures. For example, it is plausible that the spectrum could include a number of light Higgsinos. These can lead to squarks decaying via a cascade, potentially weakening our search signal, but these can give other signatures involving photons or leptons, as pointed out in \cite{Kribs:2012gx}. Our assumption of a light LSP may also not hold, and if the LSP mass is instead a significant fraction of the squark or gluino mass the sensitivity could be significantly weakened \cite{Kribs:2012gx}. Further, many well motivated Dirac gaugino models have a gravitino LSP, and if decays to this are slow on collider timescales the expected signals could be altered dramatically.

\section{Motivated parameter regions} \label{sec:model}
 
We now examine the features that models of SUSY breaking and mediation must have in order to give low energy spectra in the different parts of parameter space considered in Section \ref{sec:search}, and calculate the required EW fine tuning. We also study the properties of models that lead to a $125~\GeV$ Higgs, for different stop masses.

\subsection{Patterns of Majorana and Dirac gaugino masses}

In theories with Majorana gauginos, the gaugino masses depend primarily on the mediation mechanism, and while they change significantly when evolved to the EW scale this running is fairly independent of the other details of the model \cite{Martin:1997ns}. Often theories of gauge mediation, or string theory motivated gravity mediation, assume gauge unification leading to a prediction for the low scale gaugino masses $M_i=\alpha_i\,m_0$, in terms of a common Majorana mass $m_0$, where $M_{1,2,3}$ are the bino, wino, and gluino respectively and $\alpha_i$ is the corresponding low scale gauge coupling. Another possibility is anomaly mediation, which predicts the physical gaugino masses in terms of the gravitino mass $m_{3/2}$ and the beta functions $\beta_i$, giving $M_i\simeq\beta_i m_{3/2}/g_i $.  The sfermion masses depend strongly on both their values at the mediation scale and the details of the rest of the mass spectrum, which feeds in during running. There is a large contribution from the 
gluino mass to the soft masses of strongly coupled states in the renormalisation group (RG) equations
\beq 
\partial_t m_{\tilde{t}}^2 = - \frac{8 \alpha_s}{3 \pi} M_3^2 + ...~,
\eeq
where $t=\ln(\mu/\mu_0)$, with $\mu$ the renormalisation scale and $\mu_0$ a reference scale. As a result obtaining squarks significantly lighter than the gluino requires tuning unless the scale of mediation is extremely low (which is itself problematic). On the other hand, split-SUSY spectra in which the sfermions are significantly heavier than the gauginos are stable under the RG flow. There are various possibilities for obtaining appropriate boundary conditions for split-spectra, for example if there is an approximate R-symmetry in the SUSY breaking sector \cite{ArkaniHamed:2004fb,Giudice:2004tc,ArkaniHamed:2004yi}. Another commonly considered pattern of soft masses, known as Natural SUSY models, have low scale gluino and stop masses significantly lighter than those of the first two generation squarks \cite{Dimopoulos:1995mi,Cohen:1996vb}. The small stop and gluino masses are motivated by the large coupling of these states to the up-type Higgs, at one and two loops respectively in the RGE. Meanwhile, the 
first two generation squarks are only coupled weakly to the Higgs sector, and making them heavy alleviates collider constraints without introducing excessive tuning.

Alternatively, gauginos can get Dirac masses if extra chiral superfields $\hat{\Phi}_i =\hat{S},\,\hat{T},\,\hat{O}$, in the adjoint representation of the SM gauge group factors \uo, \suw , \suc \, respectively, are added to the theory.\footnote{Superfields are denoted with a hat.} Then Dirac masses can be generated by the operator
\beq
\int d^2\theta ~ \sqrt{2} \frac{\hat{W}_{\alpha}'\,\hat{W}^\alpha_j}{M}\hat{\Phi}_j~,
\label{eq:DGop}
\eeq 
where $\hat{W}_\alpha'$ is a hidden sector ${{\rm U(1)}}'$ spurion that gets a D-term, $M$ is the scale of supersymmetry breaking from the hidden to the visible sector, and $\hat{W}_i^\alpha$ is the visible sector gauge superfield. 
This leads to terms in the Lagrangian
\beq
\mathcal{L} \supset -m_D\lambda_j \tilde{\phi}_j - \sqrt{2} m_{D_j} (\phi_j+\phi^*_j)D_j-\frac{1}{2}D_j^2 ~,
\label{eq:dgmass}
\eeq
where $\lambda$ is the gaugino, $\phi$ is the complex scalar component of $\hat{\Phi}$ and $\tilde{\phi}$ is its fermion partner. 

The operator in Eq.~\eqref{eq:DGop} is supersoft, in contrast to a Majorana gaugino mass.
Therefore there are only threshold contributions to the sfermion masses, given by
\beq
m^2_{\tilde{f}} =\sum_i \frac{\alpha_i}{\pi} C_i\left(r\right) m_{D_i}^2 \log \left(\frac{m_{\mathrm{Re}\phi_i}^2}{m_{D_i}^2}\right) ,
\label{eq:sfmas}
\eeq
where $m_{D_i}$ is the mass of a gaugino, corresponding to the group which the sfermion is in the representation $r_i$ of, and $m_{\mathrm{Re}\phi_i}$ is the mass of the real part of the sgauge field. If only the supersoft operator in Eq.~\eqref{eq:DGop} is present, $m_{\mathrm{Re}\phi_i} = 2 m_{D_i}$ and the formula simplifies further. 
Moreover, the finiteness of Eq.~\eqref{eq:sfmas} allows for a hierarchy between the low scale squark and gluino masses to be maintained during RG flow without tuning.

The scalar adjoints can also have SUSY breaking, R-symmetry preserving, mass terms
\beq
\mathcal{L} \supset m_{\phi_i}^2 \phi_i^\dagger \phi_i + B_{\phi_i} (\phi_i \phi_i + \mathrm{h.c.})~.
\label{eq:scadmass}
\eeq
The first term of Eq.~\eqref{eq:scadmass} is actually required because otherwise the imaginary part of $\phi_i$ would be massless. However, the $B_{\phi_i}$ term splits the real and imaginary components of the adjoint scalar masses, and originates from the operator 
\beq
\int d^2\theta~ \sqrt{2} \frac{\hat{W}_{\alpha}'\,\hat{W}^{\alpha'}}{M}\hat{\Phi}^2_j~.
\eeq
From Eqs.~\eqref{eq:dgmass} and \eqref{eq:scadmass} we have
\beq
m_{\mathrm{Re}\phi_i}^2 = 4 m_{D_i}^2 + m_{\phi_i}^2+B_{\phi_i}~, \qquad \qquad  m_{\mathrm{Im}\phi_i}^2 = m_{\phi_i}^2-B_{\phi_i}~.
\eeq
Notably, the first term of Eq.~\eqref{eq:scadmass} is not supersoft, and the non-holomorphic adjoint masses $m_{\phi_i}^2$
contribute at two loops to the $\beta$ functions for the sfermion masses.

In gauge mediated models of Dirac gauginos, couplings of the adjoint fields to messengers in the superpotential lead to the soft masses of  Eqs.~\eqref{eq:dgmass} and \eqref{eq:scadmass} \cite{Amigo:2008rc,Carpenter:2010as,Benakli:2010gi}. If the messengers are charged under a ${{\rm U(1)}}'$, gaugino masses, and $B_{\phi_i}$ are generated at one loop. Avoiding a tachyonic adjoint scalar therefore requires $m_{\phi_i}^2$ to be large, which is problematic since it is generated at two loops. With many messengers, positive masses for the real and imaginary components of the scalar adjoint are possible \cite{Benakli:2008pg}. 
However, the contribution to the sfermion masses from the RG flow may then dominate the finite contribution in Eq.~\eqref{eq:sfmas} leading to problems with tachyonic states, and potentially requiring extra tuning to obtain viable spectra \cite{Arvanitaki:2013yja}. Additional operators, that give positive contributions to $m_{\mathrm{Re}\phi_i}^2$ and $m_{\mathrm{Im}\phi_i}^2$, may alleviate the problem leading to masses for the real and imaginary components of the scalar adjoint of the same order as the gluino mass \cite{Csaki:2013fla,Carpenter:2015mna,Nelson:2015cea}.
Alternatively, it is possible to forbid the operator that produces $B_{\phi_i}$ if the gauginos themselves are associated to a spontaneously broken global symmetry \cite{Alves:2015kia,Alves:2015bba}.

On the other hand, Dirac gauginos can make some aspects of model building more straightforward. For a generic hidden sector an R-symmetry is a necessary condition for supersymmetry breaking \cite{Nelson:1993nf,Shih:2007av}. 
However, Majorana gaugino masses break the R-symmetry, and it is often problematic to generate large enough gaugino masses relative to the squark masses \cite{Abel:2009ze}. This is not an issue in models with Dirac gaugino masses, since these respect the R-symmetry, and realisations of gauge mediation for example from a strongly coupled SUSY QCD like sector are possible \cite{Abel:2011dc}.

The phenomenology of Dirac gaugino models depends on the expected ratios of the scalar masses to gluinos and between the gauginos. If scalar masses are dominantly produced by the supersoft operators of Eq.~\eqref{eq:DGop}, the gluinos are significantly heavier than the squarks, with $m_{D_3} \sim (5 \div 10)\, m_{\tilde{q}}$. However more complete models can alter this minimal picture and lead to squark masses comparable to the gluino mass \cite{Carpenter:2010as}. Gauge unification is not automatic in minimal Dirac models, and if it is not imposed the ratio of gaugino masses depends on the details of the SUSY breaking and mediation sectors. It is also possible that adjoints are present only for the ${\rm SU}(3)$ group, while the wino and bino have Majorana masses, which would allow for large differences in the masses, for example if the theory has an approximate R-symmetry (options include the possibilities that the gluino could have both a Majorana and Dirac mass \cite{Ding:2015wma}, or that Dirac masses 
could 
be 
generated by an F-term \cite{Martin:2015eca}).

Unification is possible if additional fields are added, which together with the $\hat{S}$, $\hat{T}$ and $\hat{O}$ adjoints form a complete representation of a unified group: the two simplest choices being ${\rm SU(5)}$ and ${\rm SU(3)}^3$ \cite{Fox:2002bu,Benakli:2014cia}. This fixes the ratio between the gaugino masses \cite{Fox:2002bu}, and in many models the bino and wino are typically a factor of a few lighter than the gluino, often with a right handed slepton lightest supersymmetric particle (LSP).  Another possibility to achieve gauge coupling unification is to add extra states in incomplete GUT multiplets, with masses between the unification and the EW scale.\footnote{These extra fields could even play the role of gauge mediation messengers.} In some scenarios this leads to a ratio between gaugino masses given by $m_{D_i}/m_{D_j}\sim g_i/g_j$ \cite{Benakli:2010gi}.

Natural SUSY spectra with the first two generation squarks heavy can also be realised in Dirac models. For example, \cite{Benakli:2014cia} study a model with gauge unification, and stops lighter than the first two generation squarks by a factor of about $5$. The low scale  gaugino masses are
\beq
m_{D_1}/m_{D_0}:m_{D_2}/m_{D_0}:m_{D_3}/m_{D_0}\simeq0.22:0.9:3.5 \,,
\eeq
where $m_{D_0}$ is the common gaugino mass at the GUT scale. The physical stop masses are $m_{\tilde{t}}^2 \simeq 0.2 \,m_0^2+ 0.6 \,m_{D_0}^2 $, where $m_0$ is a common first two generations squark mass at the GUT scale, while the physical first two generations squark masses are approximately $m_{\tilde{q}_{1,2}}^2 \simeq 0.9\, m_0^2+ 0.6 \,m_{D_0}^2 $.

Finally, split SUSY models with Dirac gauginos are possible \cite{Fox:2014moa}. In these models, sfermions and Dirac gauginos are very heavy and the only phenomenologically viable states are the pseudo-Dirac Higgsinos, with mass around $1~\TeV$. Alternatively, the bino can be a Majorana fermion, lighter than the other Dirac gauginos, and it can generate a splitting of the pseudo-Dirac Higgsino into Majorana states. 

To summarise, 
in their simplest implementations models with Dirac gauginos lead to spectra with relatively heavy gluinos compared to the squarks without tuning, typically with gluino masses about five times larger. Meanwhile Majorana models cannot have gluinos significantly heavier than the squarks without additional tuning.  
However, minimal theories of Dirac gauginos have  problems with tachyonic states, and solving this can lead to squark masses comparable to that of the gluino.  Natural SUSY models, with the first two generation sfermions heavier than the gluino and the stops are also possible in both Majorana and Dirac models, as are split SUSY models.

\subsection{Higgs sector}   

In the MSSM, the tree level mass of the lightest neutral Higgs is constrained to be below the $Z$ boson mass and enhancing it to the observed value \cite{Aad:2012tfa,Chatrchyan:2012xdj} requires large radiative corrections from the stops. In a model with gauginos and first generation squarks above $2\,\TeV$ and light Higgsinos, the lightest stop masses allowed are about $1.7\,\TeV$ assuming maximal stop mixing \cite{Vega:2015fna}. However, if the $A$ terms are small, large stop masses of about $10\,\TeV$  are needed. 
Extra contributions to the Higgs quartic self coupling can relax the need for large radiative corrections. This happens in the NMSSM, which has a term $\lambda_S \hat{S} \hat{H}_u \hat{H}_d$ in the superpotential \cite{Ellwanger:2009dp}. In this case very light stops are possible. For $\lambda_S >0.7$ the stops can be as light as $500~\GeV$, although in some parameter ranges this leads to $\lambda_S$ running non-perturbative at an intermediate scale \cite{Hall:2011aa,Hardy:2012ef}. 

In Dirac gaugino models, if the operator in Eq.~\eqref{eq:DGop} is the only source of supersymmetry breaking, the equations of motion set the D-terms for the SM gauge interactions $D_i\equiv0$. Consequently, the tree level Higgs quartic, and its tree level mass, vanish identically. However, this is no longer the case if the terms in Eq.~\eqref{eq:scadmass} are present, and the suppression can be reduced if the soft masses for the singlet $S$ and the triplet $T$ are large enough. The suppression is also ameliorated if the gaugino masses are a mix of Dirac and Majorana. 
Models with entirely Dirac gaugino masses have an R-symmetry in the gauge sector. This may or may not be respected by the Higgs sector, leading to different ways to raise the Higgs mass to $125~\GeV$.\footnote{In both the cases with preserved or broken R-symmetry in the Higgs sector, the new couplings break custodial symmetry, potentially leading to large contributions to the EW precision observables. 
In particular the $\rho$ parameter leads an upper bound on the triplet vacuum expectation value of $|v_T|\lesssim4~\GeV$.}

If the R-symmetry is broken in the Higgs sector, there can be couplings between the singlet or the triplet adjoint and the Higgs
\beq
W\supset \lambda_S \hat{S} \hat{H}_u\cdot \hat{H}_d + \lambda_T \hat{H}_d\cdot \hat{T} \hat{H}_u ~.
\label{eq:WhRbr}
\eeq
These enhance the tree level Higgs mass at small $\tan\beta$ by an NMSSM-like term proportional to the couplings $\lambda_{S,T}$, competing with the suppression of the D-term Higgs quartic \cite{Benakli:2012cy}. A radiative contribution from the stops is still typically needed, though stop masses below $1\,\TeV$ may be sufficient depending on the value of the other parameters. Dirac gluino and wino masses can be large without disrupting this conclusion, but the Dirac bino is bounded to be below a few hundred GeV because it mixes dangerously with the lightest Higgs, reducing its mass. As a side effect, this mixing will lead to a Majorana neutralino.

 On the other hand, if the R-symmetry is preserved in the Higgs sector, extra inert Higgs-like doublets $R_{u,d}$ must be introduced to obtain a viable model, since the standard $\mu$ term is forbidden.\footnote{Models of Dirac gaugino without the $\mu$ term \cite{Nelson:2002ca} or where the $\mu$ term is generated as in the NMSSM are also possible \cite{Benakli:2011kz}.} The extra doublets can couple through R-symmetric $\mu$-type  and trilinear terms
 \bea
 W &\supset& \mu_u\, \hat{R}_u \cdot \hat{H}_u+\mu_d\, \hat{R}_d\cdot \hat{H}_d+\lambda_u\, \hat{S}\, \hat{R}_u\cdot\hat{H}_u \nonumber\\
 &+& \lambda_d\, \hat{S}\, \hat{R}_d\cdot\hat{H}_d + \Lambda_u\,  \hat{R}_u\cdot\hat{T}\, \hat{H}_u + \Lambda_d\, \hat{R}_d\cdot\hat{T}\, \hat{H}_d~.
 \label{eq:WhRsymm}
 \eea
However, the extra couplings $\lambda_{u,d}$ and $\Lambda_{u,d}$ do not alleviate the depletion of the D-terms.
For example, in the limit where $\lambda=\lambda_d=-\lambda_u$, $\Lambda=\Lambda_u=\Lambda_d$, $\mu_u=\mu_d=\mu$, and the vacuum expectation values of the singlet and triplet scalar adjoints $v_S\simeq v_T \simeq 0$, the tree level Higgs mass is
\beq
m_{h,\mathrm{tree}}^2=m_Z^2 \cos^22\beta - v^2 \left( \frac{(g_1m_{D_1}-\sqrt{2} \lambda \mu)^2}{4 (m_{D_1})^2 + m_S^2}+\frac{(g_2 m_{D_2}+ \Lambda \mu)^2}{4 (m_{D_2})^2 + m_T^2} \right)\cos^22\beta~,
\label{eq:mhRsym}
\eeq
where $m_{D_1}$ and $m_{D_2}$ are the Dirac gaugino masses and $m_S$ and $m_T$ are the masses of the real parts of the scalar adjoints before EW symmetry breaking.
Therefore the tree level upper bound is even stronger than in the MSSM, 
and radiative corrections are vital.
At one loop the most relevant corrections come from four powers of the couplings $\lambda$ and $\Lambda$. The leading contribution is~\footnote{In the limit where $\lambda=\lambda_d=-\lambda_u$, $\Lambda=\Lambda_u=\Lambda_d$, $\mu_u=\mu_d=\mu$ and $v_S\simeq v_T \simeq 0$.} (see also \cite{Bertuzzo:2014bwa,Diessner:2014ksa})
\begin{eqnarray}
\Delta m_h^2 &=& \frac{2 v^2}{16 \pi^2}\biggl[ \frac{\Lambda^2 \lambda^2}{2} + \frac{4 \lambda^4 + 4 \lambda^2 \Lambda^2 + 5 \Lambda}{8}\log\frac{m_{R_u}^2}{Q^2}\nonumber\\
&+&\left(  \frac{\lambda^4}{2}- \frac{\lambda^2\Lambda^2}{2}\frac{m_S^2}{m_T^2-m_S^2} \right)\log\frac{m_S^2}{Q^2} \nonumber\\
&+&\left( \frac{5}{8}\lambda^4+ \frac{\lambda^2\Lambda^2}{2}\frac{m_T^2}{m_T^2-m_S^2}        \right)\log\frac{m_T^2}{Q^2}\nonumber\\
&-&\left(      \frac{5}{4}\lambda^4- \lambda^2\Lambda^2 \frac{m_{D_2}^2}{m_{D_1}^2-m_{D_2}^2}     \right)\log\frac{m_{D_2}^2}{Q^2}\nonumber\\
&-&\left(    \lambda^4+ \lambda^2\Lambda^2 \frac{m_{D_1}^2}{m_{D_1}^2-m_{D_2}^2}      \right)\log\frac{m_{D_1}^2}{Q^2}\biggr]~.
\label{eq:mh1-loop}
\end{eqnarray}
The form of these is analogous to those from the stop and if $|\lambda|\sim|\Lambda|\sim 1$ the new terms can give a significant contribution.\footnote{However the condition $|\lambda|\sim|\Lambda|\sim y_t$ may lead to a loss of perturbativity at low scales \cite{Bertuzzo:2014bwa}.} For large couplings, $\Lambda_u=\Lambda_d\sim 1\sim -\lambda_u=-\lambda_d$, and light Higgsinos with mass around $300~\GeV$, it is possible to obtain the correct  Higgs mass for stops as light as $300~\GeV$ (not necessarily ruled out by the LHC if they have a compressed spectrum). A beneficial feature is that the adjoint fields do not reduce the Higgs mass at two loop, as the stop contribution proportional to $\alpha_s$ does. 

In conclusion, the physical Higgs mass leads to significant constraints on the form of viable models, especially in theories with Dirac gauginos. However, for all gluino and stop masses not ruled out by the LHC, models exist in which a $125~\GeV$ Higgs is possible. Therefore none of the parameter space of discoverable strongly interacting states studied in Section \ref{sec:search} is directly excluded, although in R-symmetric Dirac models with light stops fairly large dimensionless coupling constants are required.

\subsection{Fine tuning}

The fine tuning of a given low energy spectrum is only well defined once a UV complete theory is specified, such that the underlying parameters that can be varied are known. For example, there may be extra tuning hidden in the UV theory, or alternatively the true tuning might be reduced by particular correlations between parameters that from the low energy perspective appear independent.\footnote{The latter however raises concerns about whether assuming a UV model with such helpful correlations is itself an additional tuning.} Despite these caveats, it is interesting to make some naive estimates of the tuning in the regions of parameter space probed by the LHC and future colliders.

In the MSSM, at large $\tan \beta$, the EW scale is fixed by the relation
\beq \label{eq:mz}
M_Z \simeq -2\left(m_{hu}^2 + \left|\mu\right|^2\right) ~,
\eeq
where $m_{hu}^2$ is the soft mass squared for the up type Higgs, and the physical Higgs mass $m_{h0}$ is equal to $M_Z$ at tree level, and increased by radiative corrections.

If the mass of the stops is significantly above the EW scale there are large radiative corrections to the soft mass squared of the up type Higgs through a term in the RGEs proportional to the top quark Yukawa. Similarly, a heavy gluino leads to a large contribution to $m_{hu}^2$ at two loops,  through the stop. As a result, obtaining a low EW scale requires an unnatural cancellation of terms in Eq.~\eqref{eq:mz}, which can be quantified by the fine tuning
\beq
\Delta^{-1} = \frac{\partial \log m_{h0}^2}{\partial \log p_i} ~,
\eeq
where $p_i$ are the UV parameters of the theory that can be varied independently.

Because of the large coupling in the RGE, even for low scale mediation the low scale stop mass typically ends up close to the gluino mass. The LHC has already set strong bounds on the gluino mass. Consequently, regions of parameter space with $2\,\TeV$ stops, heavy enough to produce a $125\,\GeV$ Higgs, are typically no more tuned than those with lighter stops. Models with very light stops in special parts of parameter space that are not excluded by searches, for example close to the top mass, may even be more tuned, due to the extra cancellations needed to keep a second scalar light \cite{Hardy:2013ywa}.

In Natural SUSY models the first two generation squarks cannot be made arbitrarily heavy without introducing further tuning. Their soft masses $m_{1,2}^2$ (which are assumed close to degenerate) feed into the RGE for the stop mass squared at two loops, driving it towards tachyonic values through a term in the RGE
\beq
\frac{d m_{\tilde{t}}^2}{dt} \supset \frac{8 \alpha_3^2}{3 \pi^2} m_{1,2}^2~.
\eeq
Depending on the mediation scale it is possible to raise the first two generation squark masses a factor of roughly $10$ to $20$ above the gluino mass (i.e. the typical mass of the stop) without making the tuning of the theory worse. In extended models, for example the NMSSM, the Higgs potential is modified and  Eq.~\eqref{eq:mz} does not hold. In some circumstances, for example if there is a large coupling $\lambda \hat{S} \hat{H}_u \hat{H}_d$ to a singlet $\hat{S}$, this can reduce the sensitivity of the EW scale to the Higgs soft mass squared parameter. However, in this case the model involves extra tuning from the requirement that the Higgs properties closely resemble those of the SM Higgs \cite{Arvanitaki:2011ck}. 

In models of Dirac gauginos, the dependence of the EW scale on the up type Higgs mass remains close to that of Eq.~\eqref{eq:mz} \cite{Bertuzzo:2014bwa}. There is a tuning from the gluino mass associated to the finite contribution to the stop mass of Eq.~\eqref{eq:sfmas}, which feeds into the up type Higgs mass. However, this contribution is enhanced only by two relatively small logarithms $\log m_{Re\left(A\right)}^2/m_{\tilde{g}}^2 \times \log m_{\tilde{g}}^2/m_{\tilde{t}}^2$, unlike the MSSM where the gluino tuning varies with the mediation scale as $\log^2 m_{med}^2/m_{\tilde{t}}^2$. This raises the hopes that relatively heavy gluinos may be possible without introducing excessive tuning.\footnote{Models based on Scherk-Schwarz SUSY breaking, which can have very small EW tuning, also feature Dirac gauginos \cite{Dimopoulos:2014aua,Garcia:2015sfa}.} There is also a finite contribution to the Higgs soft mass from the wino and bino (assuming that these also have dominantly Dirac masses). For 
ratios of Dirac gaugino masses coming from typical models this leads to less tuning than that from the gluino.

Additionally, a soft mass for the imaginary part of the adjoint from Eq.~\eqref{eq:scadmass} contributes to the running of the squark masses at two loops
\beq
 \frac{d m_{\tilde{t}}^2}{dt} \supset \frac{2 \alpha_3^2}{\pi^2} m_{\phi3}^2~.
\eeq
For large sgluon masses this can be important, especially since the additional matter in Dirac models results $\alpha_3$ being larger at high scales than in the MSSM. 

To study the fine tuning as a function of the gluino and stop masses in models with a viable low energy spectrum, we fix $m_{\phi 3}$ at the mediation scale such that the low scale mass of the imaginary part of the sgluon is equal to the gluino mass. There may be additional tuning to achieve this in an actual SUSY breaking and mediation mechanism, since as discussed it is often a loop factor too large. However we do not attempt to quantify this in our measure of tuning. While it is possible that the stop mass is determined solely by the gluino mass and the negative RG contribution from the sgluon, such a setup does not allow for squark masses comparable to the gluino mass at a low scale (even allowing the mediation scale to vary). Instead we allow an extra stop soft mass generated directly at the mediation scale, opening up the low 
energy parameter space. Such a mass is not supersoft and gives a logarithmically divergent contribution to the Higgs mass squared parameter. However, in parameter ranges where the stop mass is dominantly generated by the gluino multiplet the direct stop soft mass is small by construction, so does not make the tuning significantly worse.

\begin{figure}
\begin{center} 
 \includegraphics[width=0.47\textwidth]{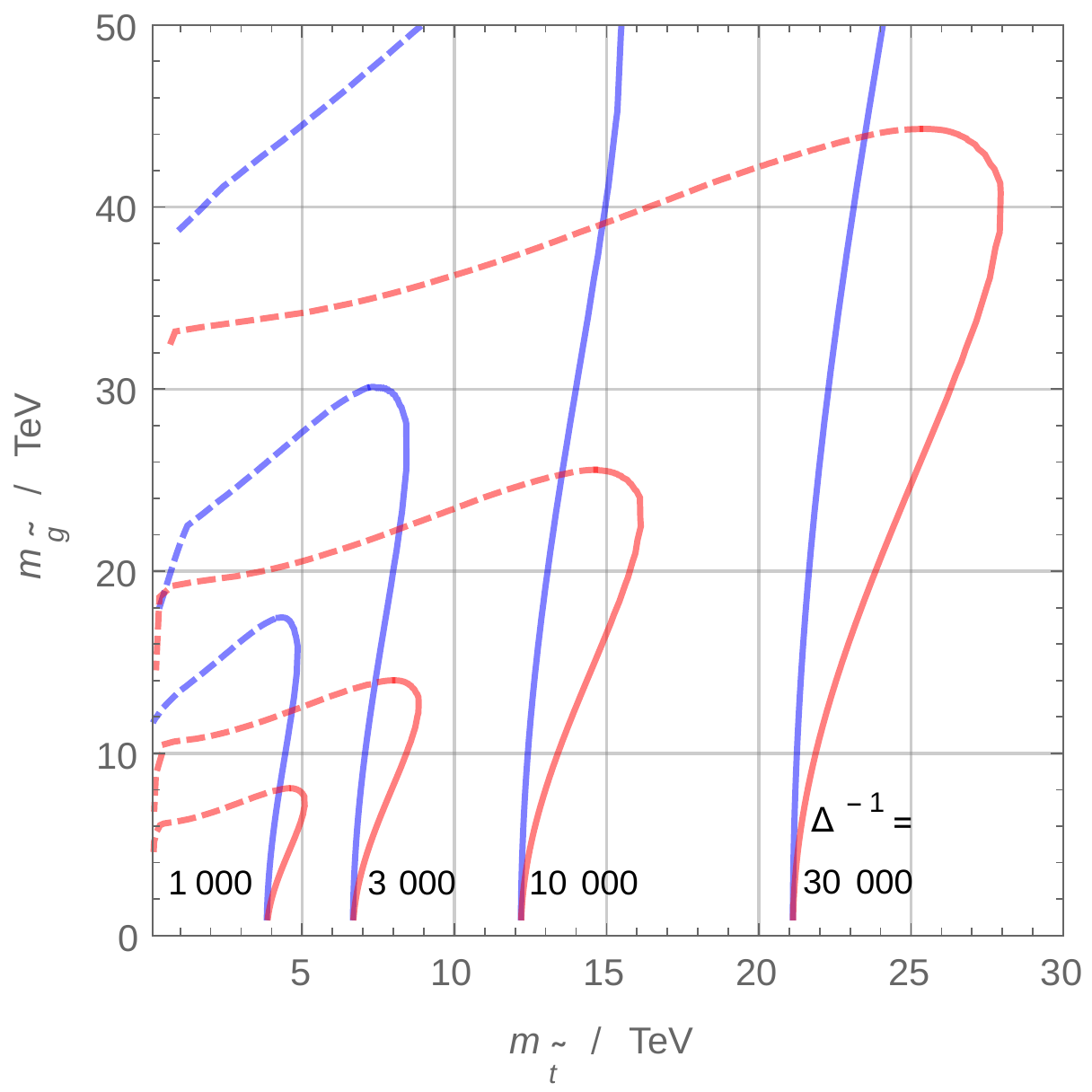}
 \qquad
 \includegraphics[width=0.47\textwidth]{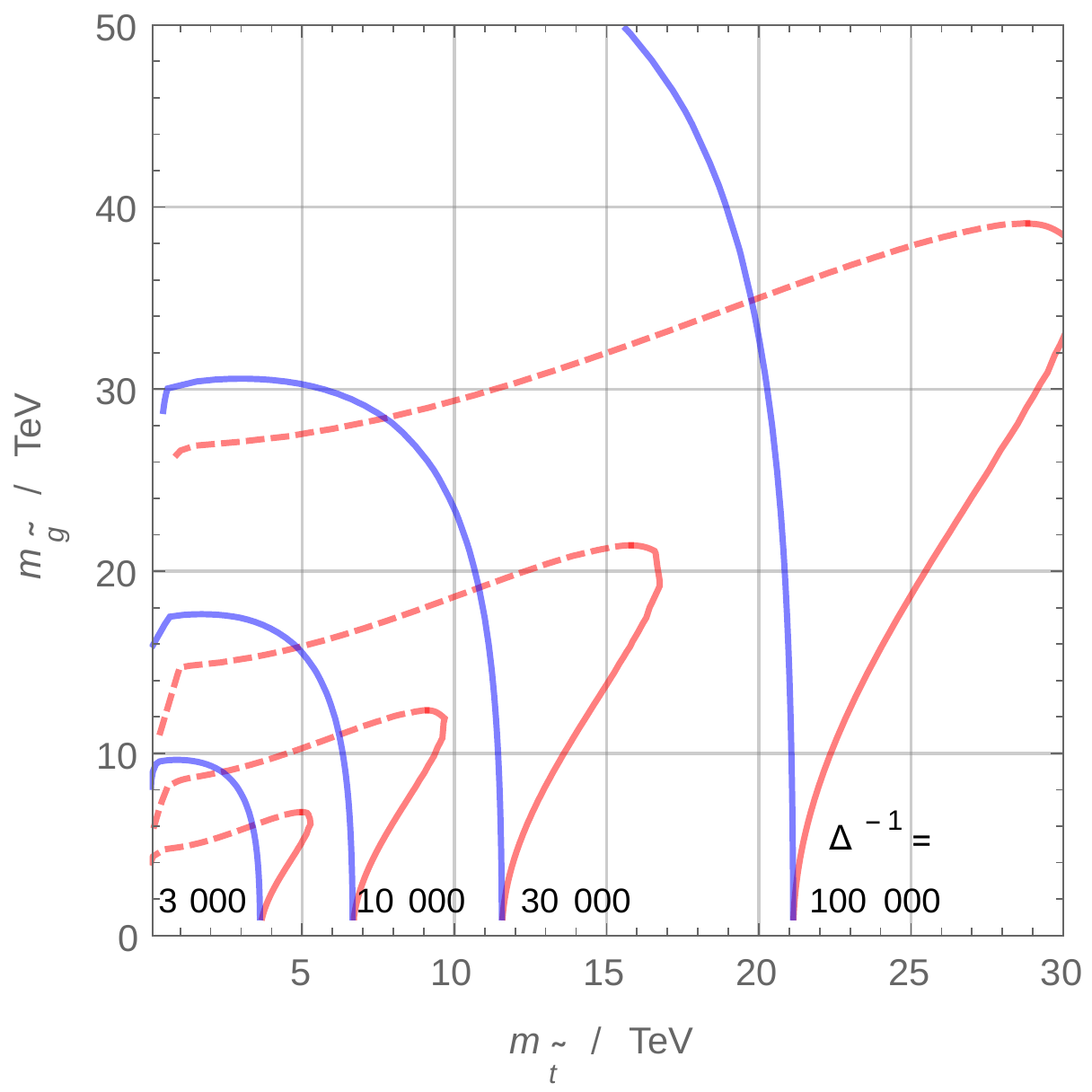}
\caption{{\bf \emph{Left:}} Contours of fine tuning for Dirac gluino models (blue) and Majorana models (red) assuming a mediation scale of $10^6\,\GeV$. The contour labels apply to both the Dirac and Majorana models, whose tuning coincides in the limit of small gluino mass. In regions with a dashed line the stop mass at the mediation scale must be tachyonic (assuming universal squark masses). {\bf \emph{Right:}} The same plot for a mediation scale of $10^{16}\,\GeV$.}\label{fig:ftz}
\end{center} 
\end{figure}

Under these assumptions we plot the fine tuning for Majorana and Dirac models in Fig.~\ref{fig:ftz} as a function of the low scale gluino stop masses. We add the individual tunings in quadrature, 
\beq
\Delta = \sqrt{\Delta_{\tilde{t}}^2 + \Delta_{\tilde{g}}^2}~,
\eeq
where $\Delta_{\tilde{t}}$ and $\Delta_{\tilde{g}}$ are the tunings from the stop and gluino parameters respectively. We also distinguish between regions of parameter space where the stop mass squared parameter at the mediation scale is negative (assuming the stop mass is degenerate with the other squarks at the mediation scale). In these parts of parameter space there are possible concerns about whether the universe could be trapped in a colour-breaking vacuum at early times (although this may not be the case \cite{Kusenko:1996jn}). Because of the sgluon and stop soft masses the tuning depends on the mediation scale even in the Dirac scenario.

Comparing with the projected collider reach studied in Fig.~\ref{fig:majoranadiscovery}, we see that in scenarios with approximately universal squarks masses of the same order as the gluino mass, a 100 TeV collider can exclude models with tuning of one part in $10,000$ for both Dirac and Majorana gluinos assuming low scale mediation and up to one part in $100,000$ for high scale mediation. Under our assumptions about the tuning, in this region there is a slight improvement in tuning in Dirac gluino models, but given the possible extra model building tunings required this is certainly not significant.

In parts of parameter space where the gluino is much heavier than the stops, Dirac models are less tuned than Majorana models especially if the mediation scale is high. This reflects the supersoftness of the gluino contribution to squark masses. The boundary between dashed and solid lines for the Dirac contours in Fig.~\ref{fig:ftz} corresponds to the stops being massless at the mediation scale. If the mediation scale is low and the squark masses are approximately universal, along this line in parameter space Dirac models with tuning of approximately one part in 1,000 will be excluded. The same low scale soft masses correspond to a tuning of almost one part in 10,000 in Majorana models. Further, the collider reach is significantly stronger for Majorana models in this parameter regime allowing models that are even more tuned to be discovered or excluded.

Natural SUSY models, with squarks far heavier than the stops, can lead to weaker discovery and exclusion potential in the gluino-stop mass plane than models with universal squark masses. Such models might therefore escape being observed with a 100 TeV collider while still having not enormous tuning. However, comparing our discussion in Section \ref{sec:search} with Fig.~\ref{fig:ftz}, we see that relatively strong searches for a gluino even in the case of decoupled squarks leads to significant constraints. For low scale mediation, Majorana Natural SUSY models will be probed up to a tuning of one part in $3,000$ and Dirac models up to similar levels depending on the efficiency of searches for stops. For high scale mediation, the longer RG flow means a Majorana gaugino has even more impact. Models with tunings of one part in approximately $30,000$ will be constrained in this case, while Dirac models with tuning of one part in 10,000 will be probed.

\section{Flavour and CP violation} \label{sec:fl}

In supersymmetric models the soft mass parameters can be a new source of flavour and CP violation. Meanwhile the violation induced in the low energy theory is strongly constrained by measurements of rare SM processes, restricting the possible patterns of soft masses in theories that can be discovered by a 100 TeV collider.
The constraints are different in models of Dirac gauginos compared to the MSSM because of the absence of a chirality changing gluino mass \cite{Kribs:2007ac}. In particular, if the deviations from universal squark masses are small---known as the mass insertion approximation---and the gluino is relatively heavy such that it can be integrated out, the leading $\Delta F =2$ flavour violating operator in the MSSM comes from operators such as
\beq
\frac{1}{m_{\tilde{g}}} \tilde{d}_R^* \tilde{s}_L^* \bar{d}_R s_L ~.
\eeq
On the other hand, in Dirac gluino models this is forbidden and the leading operators are dimension 6 contributions of the form
\beq
\frac{1}{m_{\tilde{g}}^2} \tilde{d}_L \partial_{\mu} \tilde{s}_L^* \bar{d}_L \gamma^{\mu} s_L ~.
\eeq
Therefore, for a given flavour violation in the squark sector, the rate of flavour changing processes can be suppressed by a factor of up to  $p^2 / m_{\tilde{g}}^2 \sim m_{\tilde{q}}^2/ m_{\tilde{g}}^2 $, where $p$ is a typical momentum scale, in Dirac models compared to the Majorana case.  While this suppression is an attractive feature it should however be noted that it does not extend beyond the mass insertion approximation. As shown by \cite{Dudas:2013gga}, in other regimes there is parametrically only a logarithmic suppression in the Dirac case going as $\log m_{\tilde{g}}^2/m_{\tilde{q}}^2$. Numerical factors and interference between chirality preserving and violating operators can even make Dirac models more constraining in some flavour models.

Separately to these effects, Dirac gaugino models have the extra advantage that gluino masses significantly above the squark masses are not necessarily as fine tuned as in Majorana models as seen in Section~\ref{sec:model}. Therefore, parameter regions that suppress the dangerous processes may be more natural in Dirac models.\footnote{Analysis of rare flavour processes in 5-dimensional models that lead to Dirac gaugino masses have been studied in \cite{Garcia:2014lfa}.}

We focus on the $\Delta F= 2$ constraints from the kaon sector, which are typically the strongest and most model independent. Other $\Delta F = 2$ constraints arise from the B sector, and are typically weaker, although this may not be the case if particular flavour structures are implemented. $\Delta F=1$ constraints are usually weaker, although they may be important at large $\tan \beta$, and also depend on the details of the Higgs sector. These can often be ameliorated be making the Higgs sector R-symmetric; constraints arising from the process $\mu \rightarrow e$ are discussed in \cite{Fok:2010vk}, while CP violation in the form of electric dipole moments are discussed in \cite{Hisano:2006mv}. Dirac gluino models have  flavour violation from operators generated when the sgluon is integrated out, however this is typically smaller than that from the gluon sector \cite{Plehn:2008ae}.

To calculate the rate of flavour violating processes the strongly interacting superpartners must first be integrated out. 
If the gluino and squark masses are similar, then both may be integrated out at the same time, and matched onto flavour changing operators in the effective Hamiltonian
\beq \label{eq:fvet}
H^{{\rm BSM}} = \sum_{i=1}^3\left( C_i Q_i + \tilde{C}_i \tilde{Q}_i\right) + \sum_{i=4,5} C_i Q_i~,
\eeq
where the operators $Q_i$ are dimension 6, and given in \cite{Gabrielli:1995bd,Gabbiani:1996hi}. Alternatively if there is a large hierarchy between the superpartners, the heavy superpartners must be integrated out first, and then the intermediate theory run to the scale of the lightest superpartners. The procedure in this case has been described for the MSSM in \cite{Bagger:1997gg}, and the Dirac case in \cite{Blechman:2008gu}, and we follow these references. The expressions for the coefficients $C_i$ in terms of the supersymmetric model have been expressed in terms of the squark mass matrices in, for example, \cite{Dudas:2013gga}. Below we consider the form they take in two motivated flavour scenarios: the mass insertion approximation and a hierarchical squark mass spectrum. 

After matching from the supersymmetric theory, the coefficients must be run down to the QCD scale, which we take as $2\,\GeV$, in order to make contact with lattice computations. The QCD corrections involved are significant, giving $\mathcal{O}(1)$ factors in the coefficients, which translates to factors of around $2$ in the superpartner masses \cite{Bagger:1997gg}. We use the NLO QCD expressions obtained in \cite{Ciuchini:1997bw,Ciuchini:1998ix,Buras:2001ra,Kersten:2012ed}.

To compare with experimental results requires the matrix elements of the operators between $K^0$ and $\bar{K}^0$ evaluated at the hadronic scale. These are typically given in terms of expressions in the vacuum saturation approximation \cite{Ciuchini:1998ix}, corrected by numerical bag-factors obtained from the lattice. Our inputs for these are taken from \cite{Bertone:2012cu}. The real part of the matrix element gives a contribution $\Delta m_{K}^{{\rm BSM}}$ to the mass difference between the long and short lived kaons, while the imaginary part gives a contribution to the CP violating parameter $\epsilon_K^{{\rm BSM}}$
\beq
\begin{aligned}
\Delta m_{K}^{{\rm BSM}} &= 2~ {\rm Re} \bra {K^0} H^{{\rm BSM}} \ket{{K}^0} ,\\
\epsilon_K^{{\rm BSM}} \Delta m_{K}^{\rm obs} &= \frac{\kappa}{\sqrt{2}}~ {\rm Im} \bra {K^0} {H^{{\rm BSM}}	} \ket{{K}^0} ~.\label{eq:mkv}
\end{aligned}
\eeq
Although the kaon mass difference is measured very precisely  $\Delta m_K^{\rm obs} = \left(3.483 \pm 0.006 \right)\times 10^{-15}\,\GeV$ \cite{Agashe:2014kda}, the SM prediction has a large uncertainty from low energy physics $\Delta m_K^{\rm SM} = \left(3.1 \pm 1.2 \right)\times 10^{-15}\,\GeV$ \cite{Brod:2011ty}. Therefore, we simply demand that the contribution from new physics does not exceed the measured value.\footnote{Flavour violation from the squark sector could also be measured in other ways, for example through the rate of decays to tops \cite{Kribs:2009zy}.}

For the CP violating contribution, the parameter in Eq.~\eqref{eq:mkv} $\kappa = 0.923 \pm 0.006 $, and again the experimental value is known very precisely $\epsilon_K^{\rm obs} = \left(2.228\pm 0.011\right)\times 10^{-3}$. The SM prediction at NNLO is $\epsilon_K^{\rm SM}=\left(1.81 \pm0.28\right) \times 10^{-3}$ \cite{Brod:2011ty}, in slight tension with the measured value, however it is likely the corrections from higher order are significant (further recent discussion may be found in \cite{Ligeti:2016qpi}). Therefore we demand that the contribution from new physics is $\epsilon_K^{\rm BSM}\lesssim 0.7\times 10^{-3}$.

\subsection{Mass insertion approximation}
The first scenario we consider is when the squark masses are close to a universal value $m^2_{\tilde{q}}$. In this case, flavour violation in the UV theory leads to deviations from degeneracy in the squark masses. This is parameterised by the small off-diagonal elements $m_{12}^2$ in the squark mass squared matrix, when written in a basis in which the quark mass matrix is diagonal. For kaon physics the relevant deviations are the mixings between the first and second generation squarks
\beq
\delta = \frac{m^{2}_{12}}{m_{\tilde{q}}^2} ~,
\eeq
where the mixing can be between the squarks in the left handed sector indicated by $\delta^{LL}$, squarks in the right handed sector $\delta^{RR}$, or between the two sectors $\delta^{LR}$ and $\delta^{RL}$.

Once the gluino and squarks are integrated out, the flavour violation parameterised by $\delta$ generates the operators in Eq.~\eqref{eq:fvet} in the low energy effective Hamiltonian. The size of the coefficients $C_i$ can be calculated to leading order in $\delta$, which corresponds to treating the off diagonal mass matrix elements as insertions in the relevant Feynman diagrams. The results for the Majorana and Dirac cases are given in, for example, \cite{Bagger:1997gg,Kribs:2007ac}.

\begin{figure}
\begin{center}
 \includegraphics[width=0.47\textwidth]{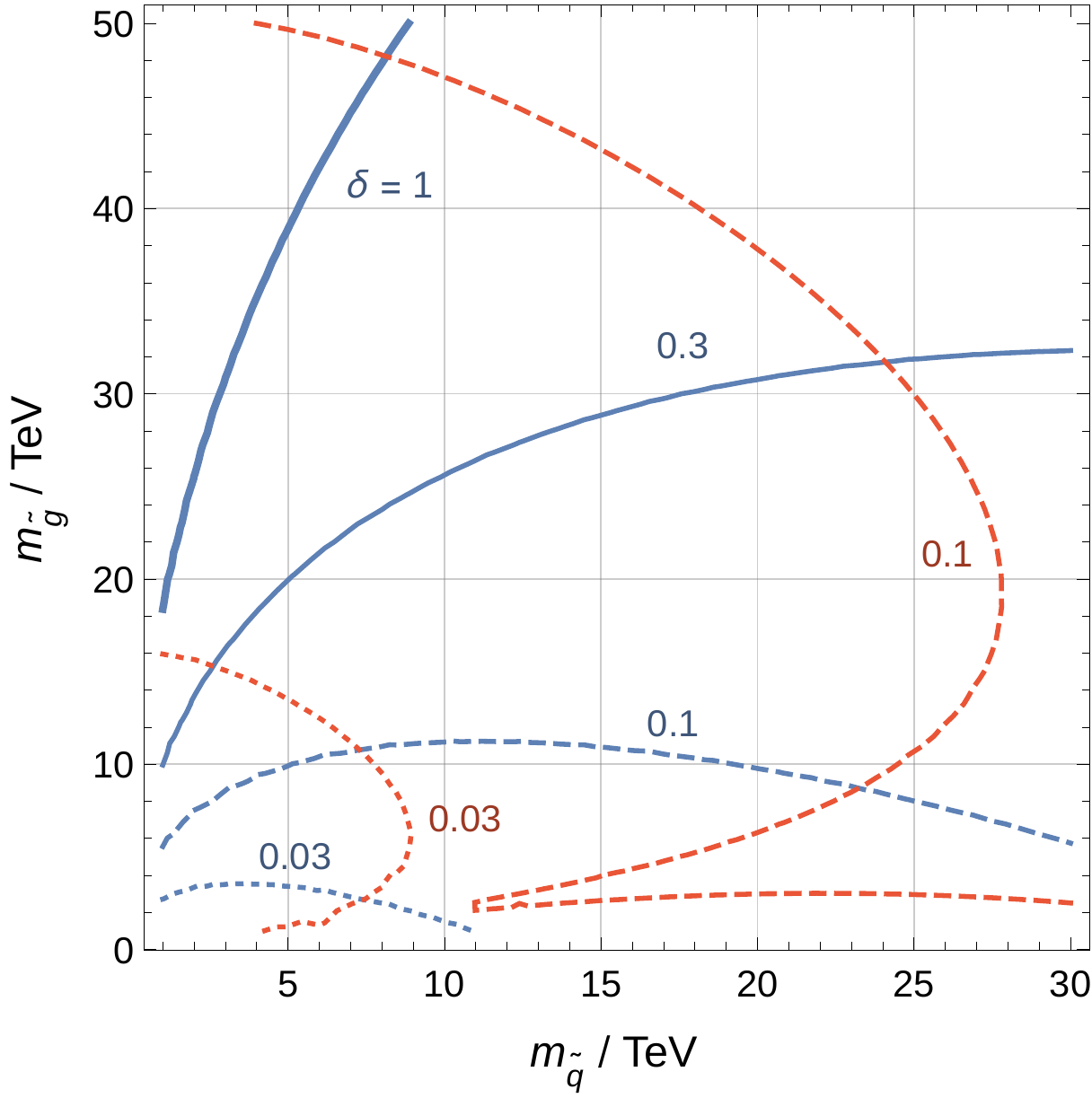}
 \qquad
 \includegraphics[width=0.47\textwidth]{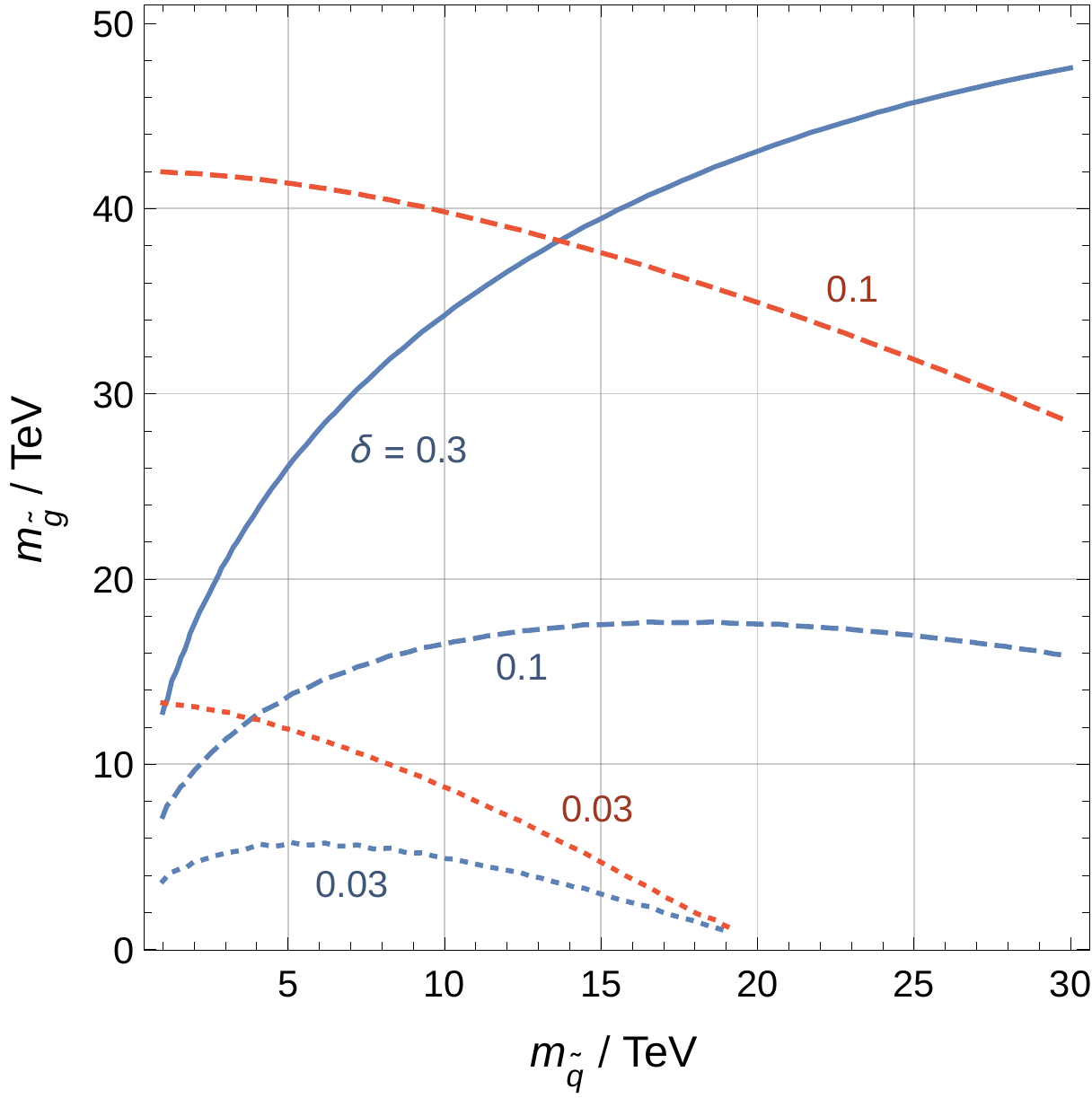}
\caption{{\bf \emph{Left:}} Contours satisfying $\Delta m_K^{{\rm BSM}} = \Delta m_K^{{\rm obs}}$ as a function of the squark and gluino masses, for different amounts of the squark flavour violation parameterised by $\delta$. This is assumed to occur in a pattern such that $\delta^{LL}=\delta^{RR}=\delta$ and $\delta^{LR}=\delta^{RL}=0$, and is shown for Dirac gluino models in blue and Majorana gluinos in red.  {\bf \emph{Right:}} The same plot if the flavour violation is such that $\delta^{LL}=\delta^{RR}=\delta^{LR}=\delta^{RL}=\delta$.}
\label{fig:fd}
\end{center}
\end{figure}

In Fig.~\ref{fig:fd} we plot the constraints obtained on superpartner masses for different values of the real part of the parameter $\delta$ in Majorana and Dirac models. The contours show the squark and gluino masses such that the bound  $\Delta m_K^{{\rm BSM}} = \Delta m_K^{{\rm obs}}$ is saturated for the given values of $\delta$.
Since, from Eq.~\eqref{eq:mkv}, the contributions to the kaon mass difference and CP violation are both proportional to the same combination $\delta^2$, the maximum acceptable imaginary part of $\delta$ as a function of the superpartner masses can be read off as ${\rm Im}~ \delta \simeq 1/23 ~{\rm Re}~\delta$. For example, the contour $\delta=1$ in Fig.~\ref{fig:fd} corresponds to an imaginary part ${\rm Im}~ \delta \simeq 1/23$ being allowed. For this contour the interpretation as a constraint on the real part of the squark flavour violation breaks down, since the mass insertion is an expansion to leading order in $\delta$. However, the interpretation in terms of the imaginary part remains accurate. 

In Fig.~\ref{fig:fd} two scenarios are shown, when $\delta^{LL}=\delta^{RR}=\delta$ and $\delta^{LR}=\delta^{RL}=0$ and also when $\delta^{LL}=\delta^{RR}=\delta^{LR}=\delta^{RL}=\delta$. The coefficients $C_4$ and $C_5$ in  Eq.~\eqref{eq:fvet} get large contributions that are proportional to $\delta^{LL} \delta^{RR}$, therefore the constraints are much weaker if the flavour violation is solely in the left or right handed sector. In these cases $\mathcal{O}(1)$ flavour 
violation is possible for squark and gluino masses less than $10~\TeV$, and $\mathcal{O}(1)$ CP violation for superpartner masses around $20~\TeV$. However, from a UV perspective, models that lead to such a flavour pattern seem relatively rare.

Dirac models are typically significantly safer in the limit where the gluino is much heavier than the squarks because of the previously mentioned suppression of the flavour changing operator by $m_{\tilde{q}}^2/m_{\tilde{g}}^2$. This can be seen analytically, for $\delta^{LL}=\delta^{RR}=\delta$ and $\delta^{LR}=\delta^{RL}=0$ the most important coefficient is $C_4$, which for Majorana and Dirac models respectively is proportional to
\bea
C_{4,m} &&\sim \frac{\alpha_s^2}{216 m_{\tilde{q}}^2}\left(504 x f_6\left(x \right) -72 \tilde{f_6}\left(x\right) \right) \delta^{LL}\delta^{RR}~, \\
C_{4,D} &&\sim \frac{\alpha_s^2}{216 m_{\tilde{q}}^2}\left(-72 \tilde{f_6}\left(x\right) \right) \delta^{LL}\delta^{RR} ~,
\eea
where $x= m_{\tilde{g}}^2/m_{\tilde{q}}^2$, and
\bea
f_{6}\left(x\right) &&= \frac{\left(-x^3+9x^2+9x-17-6\left(1+3x \right)\log x \right)}{6\left(1-x\right)^5} ~,\\
\tilde{f_{6}}\left(x\right) &&= \frac{\left(x^3+9x^2-9x-1-6x\left(1+x \right)\log x \right)}{3\left(1-x\right)^5} ~.
\eea
Therefore at large $x$ the Majorana constraint is much stronger, while at small $x$ (since the values of $f_{6}\left(x\right)$ and $\tilde{f_6}\left(x\right)$ are similar) the two scenarios have comparable bounds. This is visible in  Fig.~\ref{fig:fd}, and has been discussed in e.g.  \cite{Dudas:2013gga}.

Comparing Fig.~\ref{fig:fd} with the collider reach in Fig.~\ref{fig:majoranadiscovery}, we see that Majorana models with squarks and gluinos of similar mass must still have quite a particular flavour structure if they are to be discoverable at a 100 TeV collider, and not already ruled out by flavour constraints. Even if flavour violation is only in the left-left and right-right sectors, the parameter $\delta_{LL}$ must be less than roughly $0.1$. If left-right mixing is also allowed these parameters must be slightly smaller. The corresponding limits on CP violation are even more severe and the parameters in this sector must be a long way away from the $\mathcal{O}(1)$ values expected in many models of gravity mediation. For similar squark and gluino masses, Dirac models fare slightly better, and models with $\delta \sim 0.1$ are discoverable if there is left-right mixing. However, this still imposes strong constraints on the CP violation in such models. If the squark masses are 
significantly lighter than the gluino mass, Dirac models typically remain safer than Majorana models 
(although the later are discoverable up to larger masses in this regime). Actually in Dirac models, for a given gluino mass the flavour constrains are usually slightly weaker for lighter squarks, but the effect is only small. The general conclusion that the flavour structure must be somewhat special, and the CP violation must be even smaller, for a model to be discoverable remains in this part of parameter space.

\subsection{Natural SUSY models}

One possibility to make these limits safer is to consider Natural SUSY models. This was proposed a long time ago to help alleviate the supersymmetric flavour problem, and has been studied extensively \cite{Dine:1990jd,Dimopoulos:1995mi,Pomarol:1995xc,Cohen:1996vb,ArkaniHamed:1997ab,Giudice:2008uk,Brummer:2014yua,Mescia:2012fg}. 
The process of matching from the supersymmetric theory to Eq.~\eqref{eq:fvet} is similar to previously, although the form of the QCD corrections must be modified to account for the large splitting of the squark generations  \cite{Agashe:1998zz,Contino:1998nw,Barbieri:2010ar,Bertuzzo:2010un,Kersten:2012ed}.

In this case, the sources of flavour violation can be separated into physically distinct contributions (again following \cite{Dudas:2013gga}). The mixing of the first two generations amongst themselves will give a direct violation, which as before can be calculated in the mass insertion limit and  parameterised by the size of the off diagonal mass terms $\delta = m_{12}^2 / m_{\tilde{q}}^2$ where here $m_{\tilde{q}}$ is the average mass of the first two generation squarks. For simplicity, we consider cases with $\delta^{LL} = \delta^{RR}=\delta$ and $\delta^{RL} = \delta^{LR} =0$, but the constraints obtained are not dramatically different if $\delta^{RL} = \delta^{LR}$ are taken non-zero.

There is also a contribution to flavour violation induced by the third generation squarks because of their mixing with the first two generation squarks. This is parameterised by a number $\hat{\delta}$, which in explicit models typically satisfies $\hat{\delta} \lesssim m_{\tilde{q}}^2/ m_{\tilde{t}}^2$, where $m_{\tilde{t}}$ denotes the mass of the third generation squarks. Finally there is a contribution from a combination of the previous two effects, proportional to $\delta \hat{\delta}$. Full definitions of the mixing matrices and parameters may be found in \cite{Dudas:2013gga}.

\begin{figure}
\begin{center}
  \includegraphics[width=0.47\textwidth]{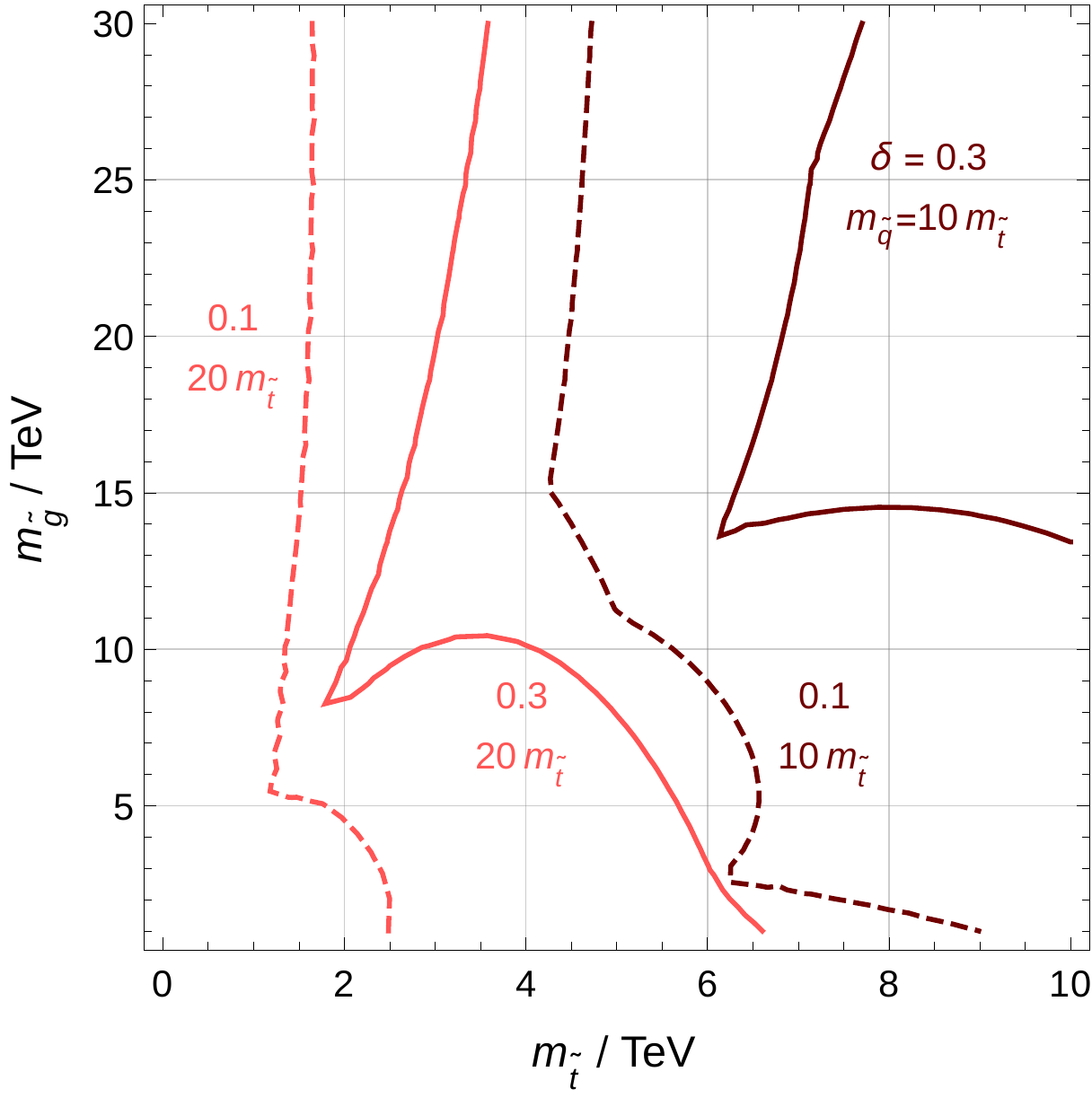}
 \qquad
 \includegraphics[width=0.47\textwidth]{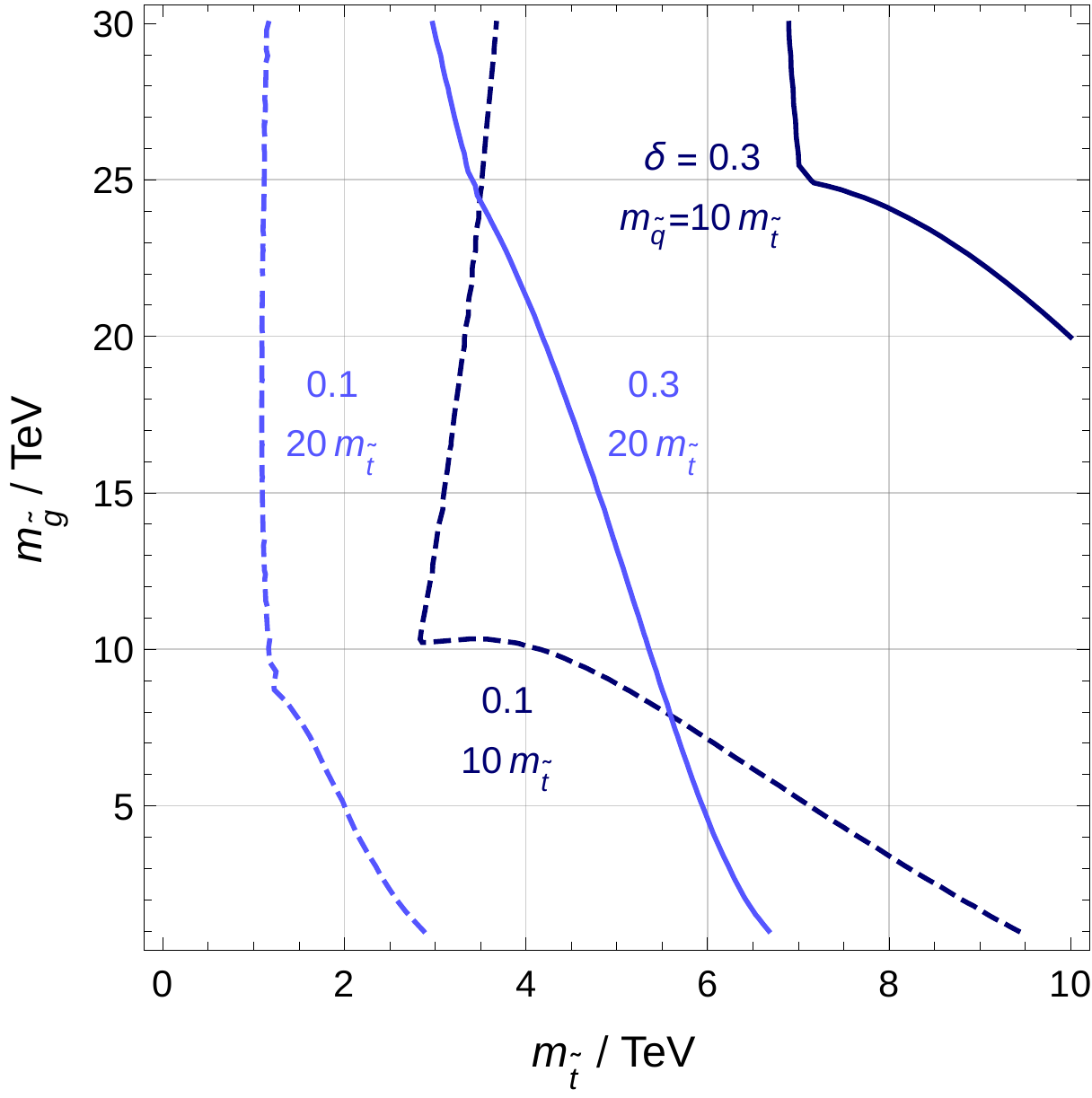}
\caption{{\bf \emph{Left:}} Contours satisfying $\Delta m_K^{{\rm BSM}} = \Delta m_K^{{\rm obs}}$ as a function of the third generation squark and gluino masses for models with Majorana gluinos and hierarchical squark masses. The first two generations squark masses are taken to be proportional to the stop masses, with $m_{\tilde{f}} = 10 m_{\tilde{t}}$ and $m_{\tilde{q}} = 20 m_{\tilde{t}}$ for dark and light lines respectively. The flavour violation induced by mixing between the third and first two generations is taken to be $\hat{\delta} = m_{\tilde{t}}^2 / m_{\tilde{f}}^2$, which is constant along contours, and arises in well motivated models. Flavour violation induced from mixing between the first two generation squarks is parameterised by $\delta= \delta^L = \delta^R$ labeled on the contours. {\bf \emph{Right:}} The same plot for models with Dirac gluino masses.} 
\label{fig:fs}
\end{center}
\end{figure}

In Fig.~\ref{fig:fs} we plot the contours of gluino and third generation squark masses that satisfy $\Delta m_K^{{\rm BSM}} = \Delta m_K^{{\rm obs}}$  for different values of ${\rm Re}~\delta$. The first two generation squark masses are taken to be a multiple of the stop masses, labelled on the contours, and the parameter  $\hat{\delta}$ is fixed to be $m_{\tilde{q}} / m_{\tilde{t}}$, constant along the contours. From a model building and fine tuning perspective, discussed in Section~\ref{sec:model}, we expect $m_{\tilde{q}} / m_{\tilde{t}} \lesssim 20$.  Further, in this plot we add the absolute values of the physically distinct contributions, that is we do not allow chance cancellations because of the relative signs of e.g. $\delta$ and $\hat{\delta}$.

From Fig.~\ref{fig:fs} it can be seen that the Dirac models lead to slightly weaker constraints in some regions, but in other regions can be even worse than Majorana models. In most parameter regions in the plot the strongest constraint comes from the first two generation mixing directly, that is the $\delta$ parameter, and there is no parametric suppression of the flavour violation induced by this in Dirac compared to Majorana models.

The expected reach for Natural SUSY models at a 100 TeV collider, discussed in Section \ref{sec:search}, allows gluinos with masses up to about $15\,\TeV$ to be discovered in both the Majorana and Dirac cases. Over the whole parameter space $m_{\tilde{g}} < 20\,\TeV$, $m_{\tilde{t}} < 15\,\TeV$, with first two generation sfermion masses 20 times the stop mass, $\delta =1$ is ruled out in both Majorana and Dirac models. Therefore, we conclude that close to, but not quite, $\mathcal{O}(1)$ flavour violation can be accommodated in well motivated, discoverable, Natural SUSY models. This corresponds to CP violating parameters no larger than order $1/100$, which remains a significant constraint.

\section{Dark matter} \label{sec:dm}

In the MSSM, if the lightest neutralino is required to be a suitable dark matter candidate the possible patterns of soft terms are strongly constrained \cite{Masiero:2004ft,ArkaniHamed:2006mb}. A pure bino LSP typically leads to a too large relic density unless it is close in mass to, and coannihilates with, the lightest slepton \cite{Ellis:1998kh}, squark \cite{Boehm:1999bj,Carena:2008mj,Cohen:2013kna} or gluino \cite{Profumo:2004wk,Feldman:2009zc,deSimone:2014pda,Harigaya:2014dwa}. Alternatively resonant s-channel annihilation through a heavy Higgs can also lead to viable scenarios. In this case the bino can be as heavy as $300~\GeV$ for stau coannihilation, $1.8\,\TeV$ for stop coannihilation, and $7.5\,\TeV$ if degenerate with a gluino. On the other hand, if the LSP is purely Higgsino, 
annihilation is dominantly into gauge bosons and 
coannihilation is often automatically present since the charged and neutral Higgsinos are usually 
almost degenerate. The 
correct relic density is obtained for Higgsino masses about $1\,\TeV$. Similarly, for a pure wino, coannihilation happens with the charged winos, and the correct relic density occurs 
for $M_2\sim 3\,\TeV$ (Sommerfeld enhancement is important in this case \cite{Hryczuk:2010zi,Hryczuk:2011tq}). The correct dark matter relic density can also be obtained from a bino-Higgsino, bino-wino (the well tempered neutralino \cite{ArkaniHamed:2006mb}), or wino-Higgsino mixed LSP. In all of these cases the mass splitting between the LSP and the NLSP is required to be around $20$ to $30~\GeV$ for coannihilation.

On the other hand, if the scale of SUSY breaking is low the gravitino can act as dark matter. The relic density is then determined by thermal production and  decays of unstable particles, and also depends on the cosmological history of the universe. Meanwhile collider signatures depend strongly on the nature and lifetime of the NLSP. If the gravitino is sufficiently heavy the NLSP appears stable at colliders (so may look like a dark matter candidate if it is a neutralino). However if the NLSP is charged and decays in the detector the charged track allows a measurement of its lifetime and can allow an estimate of the gravitino mass. Gravitinos are so weakly interacting that they cannot be detected by direct detection searches and their annihilation cross sections are so suppressed that indirect detection signals are negligible. 

The viable dark matter candidates can change dramatically if gauginos have Dirac masses. In models of gauge mediation the gravitino can be a suitable dark matter candidate. However, similarly to the MSSM the possible collider signatures are highly model dependent, so we focus on the high mediation scale case where a neutralino is the LSP. 

If there are additional chiral adjoints only for the \suc\, gauge group, while the other gauginos are Majorana and the Higgs sector is unchanged from the MSSM, viable dark matter candidates are as for the MSSM. For the purposes of a 100 TeV collider, such models simply change the relations between gluino searches and dark matter. It is interesting that large hierarchies between the gluino and other gaugino masses can easily occur in this scenario, for example due to an approximately conserved R-symmetry. Such a spectrum relaxes the links between collider searches for coloured states and dark matter, which we discuss shortly. Similarly, if chiral adjoints are present for the \suw\, and \uo\, groups, but the mediation is such that these gauginos are dominantly Majorana, the dark matter candidates are similar to the MSSM, with the extra possibility that the LSP could have a significant adjoint fermion component. New dark matter scenarios occur when the LSP 
is a combination of the MSSM-like gaugino and the adjoint fermion, and when the bino and wino in the MSSM are replaced with the corresponding adjoint fermion, leading to to dark matter candidate with either Majorana or Dirac masses \cite{Hsieh:2007wq,Harnik:2008uu,Benakli:2008pg,Belanger:2009wf,Benakli:2010gi,Benakli:2012cy,Benakli:2014cia,Goodsell:2014dia,Goodsell:2015ura}. 

On the other hand, if the theory has an unbroken R-symmetry neutralinos are a linear combination of the bino/adjoint singlet (Dirac bino), wino/adjoint triplet (Dirac wino), up and down Higgsinos, and extra so called R-Higgsinos that must be introduced. We take this scenario, called Minimal R-symmetric Supersymmetric SM (MRSSM), 
as a representative example to study Dirac gaugino masses and it is reviewed in Appendix \ref{app:MRSSM}. Another motivated possibility is that the mediation mechanism is such that all the gaugino masses are dominantly Dirac, but the Higgs sector is that of the MSSM. The phenomenology of this case qualitatively follows that of the MRSSM.
In these cases the Dirac nature of the gauginos has a large effect on direct detection searches. 

In the following we study the relationship between direct and indirect detection, relic density and collider reach for Dirac gaugino models. For our numerical results we get model inputs from the CalcHEP \cite{Belyaev:2012qa} output of SARAH \cite{Staub:2013tta}, the mass spectrum and couplings are computed at one-loop with SPheno \cite{Porod:2011nf}, and finally the relic density and direct and indirect detection rates are computed with MicrOMEGAs \cite{Belanger:2013oya}. 

\subsection{Direct detection and relic density}

Direct and indirect dark matter detection experiments can be used to place bounds on the dark matter mass. The Dirac or Majorana nature of the lightest neutralino dramatically changes the interactions probed by direct detection experiments. 
For Majorana particles the vector interaction with quarks vanishes and the neutralino--nucleon cross section is suppressed. Therefore in the MSSM the dominant process for the spin independent cross section are Higgs and squark exchange, while Z exchange contributes only to the spin dependent cross section. 
Combining the relic density constraint  \cite{Ade:2015xua} and bounds from direct detection experiments, in most models  a significant part of the parameter space is ruled out already for bino/Higgsino and Higgsino/wino LSP, although there are still viable regions \cite{Bramante:2014tba}.
Moreover, indirect detection may set limits on wino dark matter with masses between $500~\GeV$ and $3\,\TeV$ \cite{Hryczuk:2014hpa}, although there are large astrophysical uncertainties.

In contrast, in models of Dirac gauginos the vector interaction of the Z exchange can lead to a large contribution to the spin independent cross section if the dark matter has a significant Higgsino content \cite{Buckley:2013sca}. 
In general  the spin independent cross section with protons can be written as 
\beq
\sigma_{SI} = \left(\frac{m_\chi m_p}{m_\chi+m_p}\right)^2 \frac{1}{16 \pi \,m_\chi^2 m_p^2}\left( \frac{1}{4}\sum_{\mathrm{spins}}|\mathcal{M}|^2\right)~,
\eeq
where $m_\chi$ is the DM mass and $m_p$ is the proton mass.
The Z-exchange spin independent cross section with protons is given by
\begin{eqnarray}
\sigma_{SI}^p &\simeq& \left(\frac{m_\chi m_p}{m_\chi+m_p}\right)^2 \frac{1}{\pi\, m_Z^4} c_{\chi \chi Z}^2 \left( \sum_{q=u,d} c_{q\bar{q}Z} B_{qZ}^p \right)^2 \nonumber\\
&\simeq& \left(\frac{m_\chi m_p}{m_\chi+m_p}\right)^2 \frac{g^4\left(\frac{1}{2} - 2 s_W^2\right)^2}{\pi c_W^4 m_Z^4}  \left(|N_{13}^{(1)}|^2-|N_{14}^{(1)}|^2+|N_{13}^{(2)}|^2-|N_{14}^{(2)}|^2\right)^2 ~,
\end{eqnarray}
where $s_W$ ($c_W$) is the sine (cosine) of the weak mixing angle, $c_{\chi\chi Z}$ is the coupling between two DM particles and the Z boson, $g$ is the \suw\, coupling and $N_{ij}^{(1)},N_{ij}^{(2)}$ are the unitary mixing matrices that diagonalise the neutralino mass matrix, in a basis where $N_{13}$ and $N_{14}$ corresponds to the Higgsino content of the dark matter (complete definitions are in Appendix \ref{app:MRSSM}).\footnote{Depending on the different values of the parameters characterising the mass matrix, $N_{1j}^{(1,2)} \lesssim 1$. If the dark matter is a pure state, the corresponding element is $\mathcal{O}(1)$, while the others vanish.}
The squark-exchange spin independent cross section with protons, for bino dark matter, is given by
\begin{eqnarray}
\sigma_{SI}^p &\simeq&  \left(\frac{m_\chi m_p}{m_\chi+m_p}\right)^2 \frac{g_1^4}{4\, \pi\, m_{\tilde{q}}^4}\left( Y_{u_L}^2 + Y_{u_R}^2 +\frac{Y_{d_L}^2}{2} +\frac{Y_{d_R}^2}{2} \right) ~,
\end{eqnarray}
where $m_{\tilde{q}}$ is the squark mass and $Y_i$ are the hypercharges of the different quarks.\footnote{In the wino dark matter case, we have to replace $g_1\to g_2$.} Finally the Higgs-exchange spin independent cross section with protons is given by
\begin{eqnarray}
\sigma_{SI}^p &=& \left(\frac{m_\chi m_p}{m_\chi+m_p}\right)^2 \frac{c_{\chi \chi h}}{ \pi\, m_h^4}\left( \sum_q c_{q \bar{q} h} B_{q \bar{q} h}^p \right)^2 ~,
\end{eqnarray}
where $c_{\chi\chi h}$ is the coupling between DM and the Higgs boson and $c_{q \bar{q}h}$ is the coupling of the Higgs with the quarks. 
The Yukawa suppression in the Higgs contribution makes the Higgs-exchange diagram subdominant to the Z- and squark-exchange, while interference between these is only important  if their amplitudes are comparable. 
Defining 
\beq
Z_N=|N_{13}^{(1)}|^2-|N_{14}^{(1)}|^2+|N_{13}^{(2)}|^2-|N_{14}^{(2)}|^2 ~,
\label{eq:mudec}
\eeq
we have that
\bea
\sigma_{SI}^{p,Z} &\simeq& 8\times 10^{-46} \left(\frac{Z_N}{10^{-3}}\right)^2 \,\,\mathrm{cm}^2,\nonumber\\
\sigma_{SI}^{p,\tilde{q}} &\simeq&  8\times 10^{-46}\left(\frac{3.5\, \mathrm{TeV}}{m_{\tilde{q}}}\right)^4\,\,\mathrm{cm}^2 ~.
\label{eq:sigmaSI}
\eea

\begin{figure}
\begin{center}
 \includegraphics[width=0.7\textwidth]{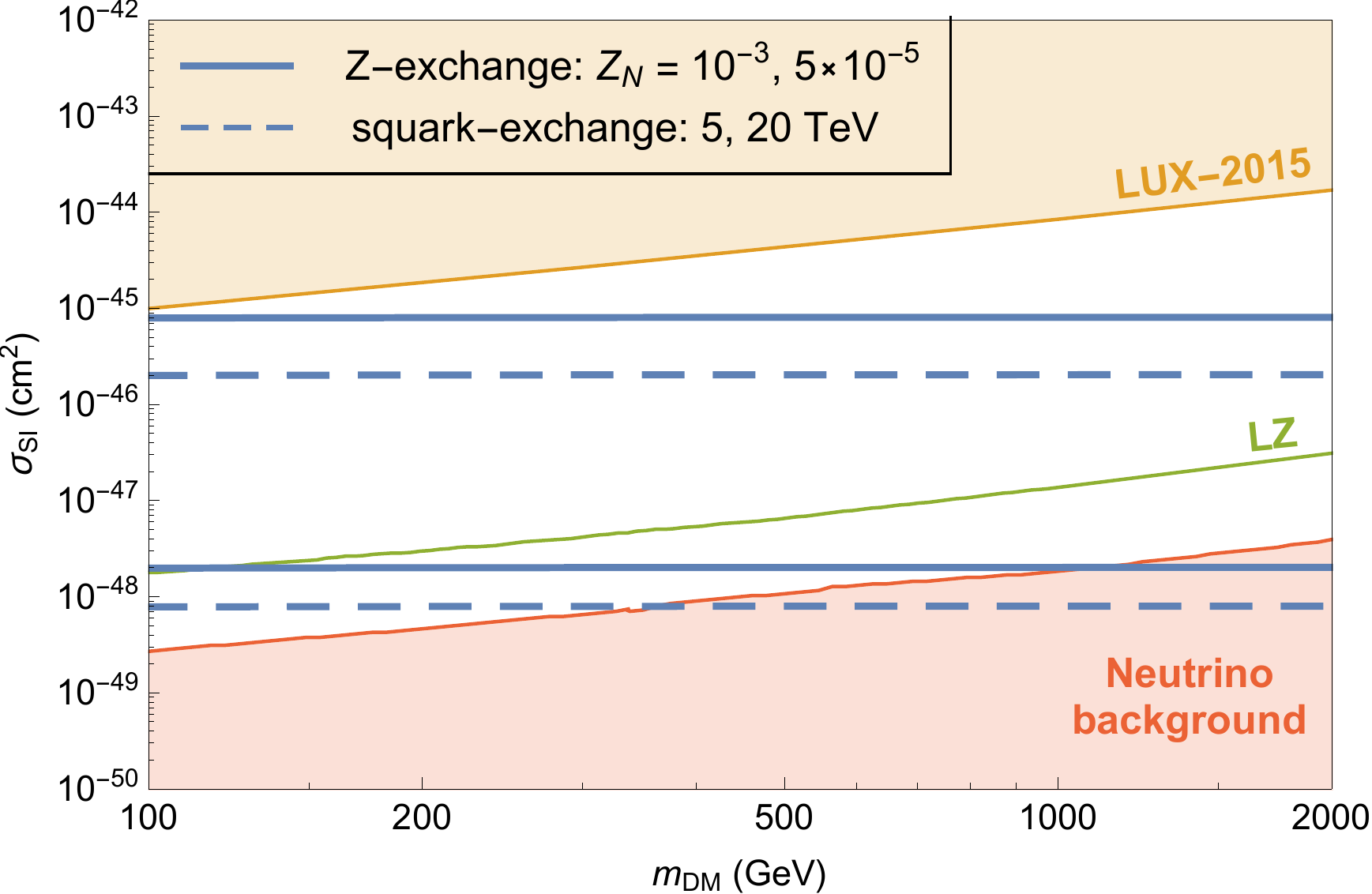}
\caption{The proton elastic scattering cross section due to Z and squark exchange processes. The two blue solid lines show the cross section for the Z-exchange for LSP Higgsino contents of $Z_N\simeq10^{-3}$  (upper curve) and $Z_N\simeq5\times10^{-5}$ (lower curve). In the limit of Eq.~\eqref{eq:mudec2} these correspond to Higgsino masses of $1.3 \,\TeV$ and $6\,\TeV$ respectively. The blue dashed lines show the cross section for squark-exchange for squark masses of $5\,\TeV$ (upper) and $20\,\TeV$ (lower), assuming a dominantly bino dark matter candidate. The orange shaded region is the actual bound from LUX, while the green curve is the projected reach of LZ. The red area denotes the neutrino background.}
\label{fig:SIgen}
\end{center}
\end{figure} 
In Fig.~\ref{fig:SIgen} we plot the different contributions to the spin independent elastic scattering cross section. The spin independent cross section for Z-exchange is shown for two benchmark values of the Higgsino content of the LSP: the upper curve has $Z_N\sim 10^{-3}$, while the lower curve  has $Z_N\sim5\times10^{-5}$. The dashed lines give the spin independent cross section for squark-exchange with squark masses $5\,\TeV$ and $20\,\TeV$, assuming bino dark matter. 
Therefore to avoid the bound from LUX  \cite{Akerib:2013tjd}, the neutralino needs a small Higgsino content\footnote{There are however blind spots, 
where a tuning of the parameters may lead to $N_{13}^{(1)}=N_{14}^{(1)}$ and $N_{13}^{(2)}=N_{14}^{(2)}$, giving rise to 
a vanishing vector current and suppressing the spin independent cross section.} and the squarks must be 
heavier than roughly $3\,\TeV$. The LZ experiment \cite{Malling:2011va,Cushman:2013zza} will be able to probe a Higgsino content of the LSP as small as $5\times10^{-5}\,\TeV$ and squarks around $20\,\TeV$. In the plot we show also the neutrino background, where direct detection experiments lose sensitivity \cite{Billard:2013qya}.

A small Higgsino content typically requires a heavy Higgsino. For example, in the limit where $M_D^B<\mu=\mu_u=\mu_d \ll M_D^W$, with $\lambda=\lambda_u=\lambda_d=\Lambda_u=\Lambda_d\sim0$ and large $\tan\beta$ we have
\beq
N_{11}^{(1)} \simeq 1- \left(\frac{g_1 v M_D^B}{4(\mu^2-(M_D^B)^2)}\right)^2, \qquad N_{12}^{(1)} \simeq N_{13}^{(1)} \simeq 0, \qquad  N_{14}^{(1)} \simeq \frac{g_1 v M_D^B}{2(\mu^2-(M_D^B)^2)} ~,
\label{eq:N_1}
\eeq
and 
\beq
N_{11}^{(2)} \simeq 1- \left(\frac{g_1 v \mu}{4(\mu^2-(M_D^B)^2)}\right)^2, \qquad N_{12}^{(2)} \simeq N_{13}^{(2)} \simeq 0, \qquad  N_{14}^{(2)} \simeq \frac{g_1 v \mu}{2(\mu^2-(M_D^B)^2)}~.
\label{eq:N_2}
\eeq
Inserting Eqs.~\eqref{eq:N_1} and \eqref{eq:N_2} into \eqref{eq:mudec} gives
\beq
Z_N \simeq \frac{g_1^2 v^2}{4} \frac{\mu^2+(M_D^B)^2}{(\mu^2 -(M_D^B)^2)^2} ~,
\label{eq:mudec2}
\eeq
and therefore Higgsinos heavier than $1\,\TeV$ are needed for bino masses around $100 \,\GeV$ to avoid the bound from LUX.  LZ may explore parameter regions corresponding to Higgsinos lighter than $4.5\,\TeV$ for $100~\GeV$ Dirac bino dark matter.

With light Higgsino scenarios excluded, the wino or the bino remain as possible dark matter candidates. A wino LSP is hard to achieve from a model building perspective, since it turns out that in most of the parameter space the lightest wino is a chargino \cite{Kribs:2008hq}. Consider, for example, the limit where $\Lambda\simeq\lambda\simeq0$, $\tan\beta\gg1$ and $M_D^B \gg \mu_d,\,\mu_u,\,M_D^W$. After the bino has been integrated out, the mass matrices for the neutralinos and the charginos are\footnote{The neutralino mass matrix is in the basis $\left(\tilde{W}^0, R_d^0, R_u^0\right), \left(\tilde{T}^0, \tilde{H}_d^0, \tilde{H}_u^0\right)$, while the chargino mass matrix is in the basis $\left(\tilde{W}^-, R_u^-\right), \left(\tilde{T}^+, \tilde{H}_u^+\right)$.} 
\beq
M_\chi=\left( \begin{array}{ccc}
m_D^W & -m_W & 0 \\
0 & \mu & 0\\
0 & 0 & \mu \end{array} \right)~,\qquad\qquad \mathrm{and} \qquad\qquad m_{\tilde{\rho}^{-}}=\left( \begin{array}{cc}
m_D^W & \sqrt{2} m_W  \\
0 & \mu \end{array} \right)~.
\eeq
The neutralino mass matrix is simplified by the Higgsino (in the third row) not mixing with the other states. The upper left $2\times2$ block has the same form as the chargino mass matrix $m_{\tilde{\rho}^{-}}$,  
but the off-diagonal element is smaller. This means that the lightest chargino is lighter than the lightest neutralino both in the Higgsino-like limit ($\mu<m_D^B,\,m_D^W$)  
and the wino limit ($m_D^W<m_D^B,\,\mu$), preventing a pure wino being a viable dark matter candidate. This feature of the MRSSM appears in large portions of parameter space for arbitrary values of $\Lambda,\,\lambda$ away from the limit considered here \cite{Kribs:2008hq}. 
\begin{figure}
\begin{center}
 \includegraphics[width=0.47\textwidth]{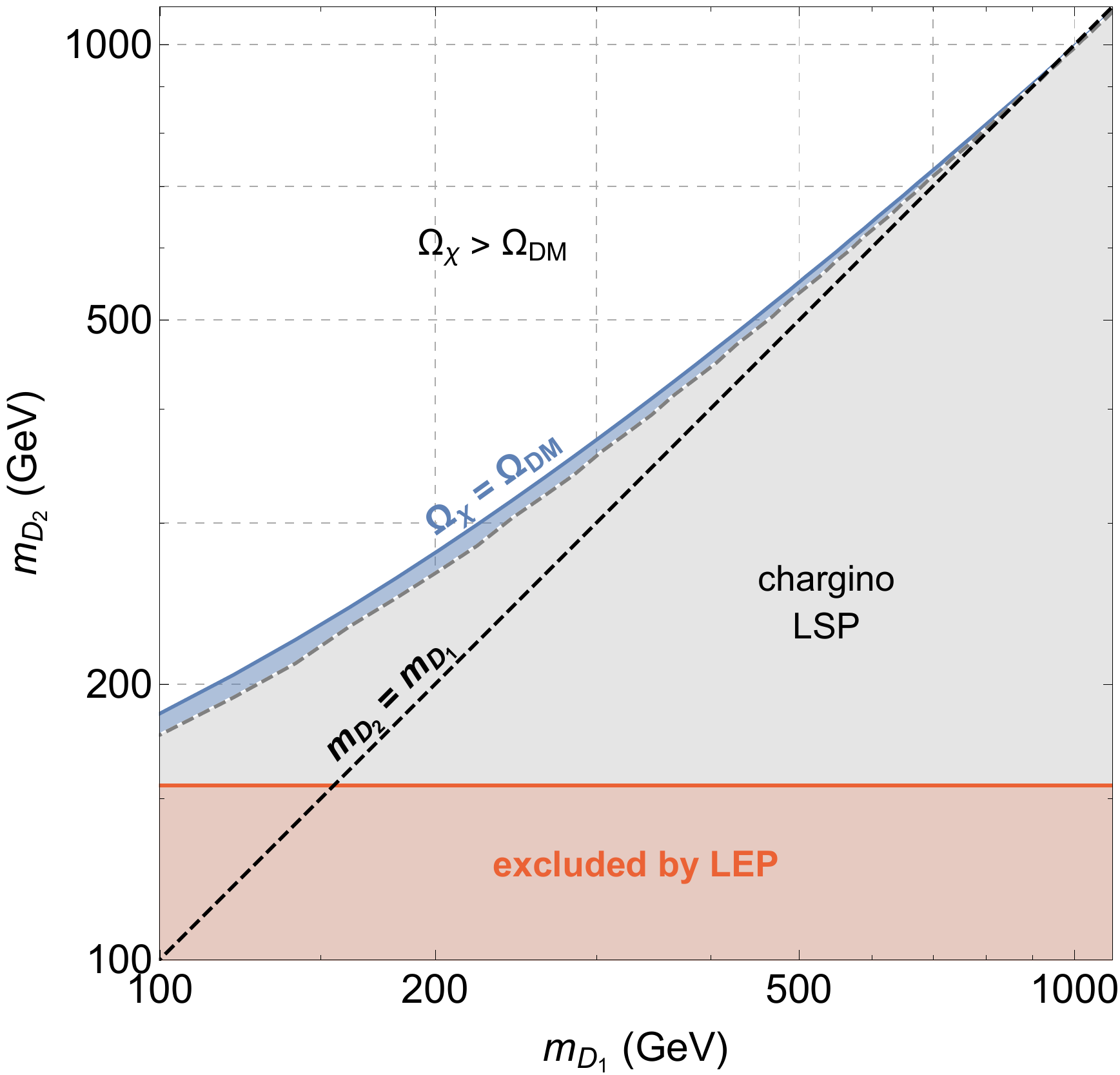}
 \qquad
  \includegraphics[width=0.47\textwidth]{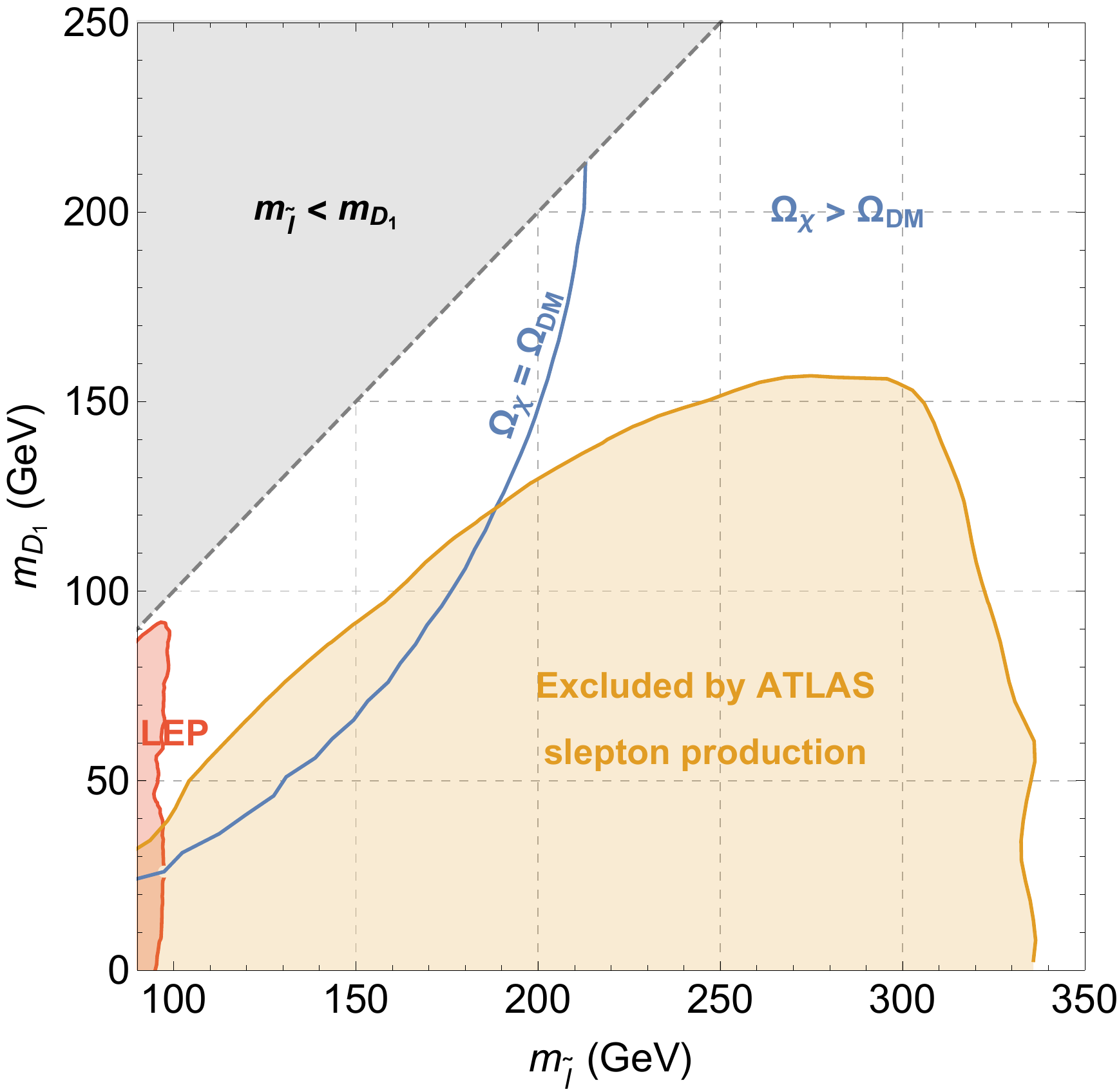}
\caption{{\bf \emph{Left:}} The constraints on Dirac bino-wino dark matter parameter space as a function of their soft masses $m_{D_1}$ and $m_{D_2}$. Models in the thin blue strip can give the correct dark matter relic density, while the relic density is too large in the white region. In the grey region the theories have a chargino LSP, and we also show the LEP constraint on light charginos \cite{LEPcha}. {\bf \emph{Right:}} The constraints on Dirac bino dark matter annihilating through sleptons, as a function of the bino mass $m_{D_1}$, and the common first two generation slepton soft mass $m_{\tilde{l}_{1,2}}$. The correct relic density is obtained along the blue contour, and the relic density is too large to the right of this. The ATLAS search for sleptons is shown in yellow and excludes a significant part of parameter space, while the constraints from LEP are shown in red.}
\label{fig:BW}  
\end{center} 
\end{figure} 

A small part of parameter space remains in which a mostly bino LSP can coannihilate with the winos to give the correct relic density.
 In Fig.~\ref{fig:BW} we plot the parameter space for a bino-wino dark matter candidate, where squarks, sleptons and Higgsinos are decoupled with masses of order $10\,\TeV$. The singlet and triplet vacuum expectaction values are $v_S\simeq0.1 ~\GeV$ and $v_T\simeq0.02 ~\GeV$, evading bounds from EW precision observables \cite{Bertuzzo:2014bwa,Diessner:2014ksa}.  We take large $\tan\beta$ to enhance the tree level Higgs mass, and the couplings $\lambda_d=-\lambda_u=\lambda\simeq-0.1$ and $\Lambda=\Lambda_d=\Lambda_u\simeq0.5$ giving a Higgs mass of approximately $125~\GeV$ over the whole parameter space.\footnote{There are other possibilities for reaching a  $125~\GeV$ Higgs mass. However, varying the values of $\Lambda$ and $\lambda$, keeping  $v_{S,T}$ fixed to have the right Higgs mass we did not find any appreciable difference in the relic density, while the spin independent cross sections  vary slightly but remain below the reach of LZ everywhere.} 
Over a large part of parameter space the LSP is a chargino, and the only viable models are in a thin strip, where a mainly bino neutralino can be the dark matter candidate. In this region the mass splitting between the bino dark matter and the mostly wino NLSP is around $20~\GeV$, similar to the MSSM bino-wino scenario.
However, in Dirac models the expected signal from indirect detection searches is reduced with respect to the MSSM.

In the pure bino case annihilation must proceed through sleptons to give the correct relic density, while avoiding direct detection constraints on squarks and Higgsinos. For pure Dirac states bino annihilation is dominated by the t-channel sfermion exchange, and the largest contribution is from right handed sleptons due to the their large hypercharge.  The annihilation is relatively slow,  $\sigma_{\mathrm{ann}} \propto m_{\chi^0}^2/m_{\tilde{l}_R}^4$, therefore the bino and the sleptons need to be close to the LEP limit and 
have similar masses to reduce the relic density sufficiently. In this case the dominant annihilation mechanism is $\chi_1^0\,\bar{\chi}_1^0 \to l_i^+ l_i^-$, and there is also  coannihilation with sleptons mainly via $\tilde{l}_i \chi_1^0\to\gamma\, l_i$ or $Z\,l_i$, and $\tilde{l}_i\tilde{l}_j^*\to \gamma\, \gamma$, $\gamma Z$ or $ZZ$.

In Fig.~\ref{fig:BW} (right) we plot the contour with the correct relic density for a bino-slepton scenario. The model parameters are the same as in the left panel, but now the first two generation of sleptons are light (we keep the stau heavy) and the wino is decoupled with a mass of order $10\,\TeV$. With lighter staus it is possible to have bino dark matter with a mass up to about $300~\GeV$. Fig.~\ref{fig:BW} (right) also shows the constraints from LEP \cite{LEP} and from ATLAS \cite{Aad:2014vma}. Although the viable parameter space is relatively small, it is easier to obtain the correct relic density than in such a scenario in the MSSM. This is because in the MSSM the dominant annihilation process is P-wave suppressed and a large enough annihilation cross section is only possible with close to degenerate bino and slepton masses. In contrast, in Dirac models with an unbroken R-symmetry the annihilation cross section into a fermion and an anti-fermion pair has a non-vanishing S-wave 
contribution even in the limit of vanishing fermion masses.\footnote{If the R-symmetry is broken the self annihilation cross section for the processes $\chi_1^0\,\chi_1^0 \to l_i^+ l_i^-$ and $\chi_2^0\,\chi_2^0 \to l_i^+ l_i^-$ are P-wave suppressed (analogously to the MSSM), while the process $\chi_1^0\,\chi_2^0 \to l_i^+ l_i^-$ has a non vanishing S-wave 
contribution (as in the MRSSM).}

The large annihilation of Dirac binos into charged lepton pairs leads to different indirect detection signatures with respect to the MSSM. Dark matter annihilation to leptonic channels does not contribute to the signal of cosmic ray antiprotons, yielding a very weak constraint. The strongest indirect detection constraints come from the annihilation to $e^+e^-$, leading to a bound $m_\chi\gtrsim 60~\GeV$ for an annihilation cross section of order $10^{-26}\,\,\mathrm{cm}^3/\mathrm{s}$ \cite{Bergstrom:2013jra}. 

Finally, small Majorana mass terms can slightly break the R-symmetry and change the behaviour of the dark matter. For example, in Split Dirac Supersymmetry models \cite{Fox:2014moa}, the relic abundance is as it is in the usual Dirac scenario, while direct and indirect detection signals may or may not be, depending on the splitting between the two Majorana states. In particular, for splittings larger than a few keV the neutralino scattering through $Z$-exchange is suppressed, as in Majorana models. In this framework it is then possible to have a pseudo-Dirac Higgsino LSP as a dark matter candidate with a mass around $1\,\TeV$.

\subsection{MSSM dark matter at colliders}

Beside direct and indirect detection experiments, information on dark matter scenarios can be obtained from collider searches. These can constrain neutralino dark matter models  through searches for the LSP, for example neutralino pair production leading to monojet signatures.\footnote{Mono-photon, -W, -Z or even -Higgs are possible \cite{Beltran:2010ww,Fox:2011pm,Gershtein:2008bf,Fox:2011fx,Bai:2012xg,Petriello:2008pu,Carpenter:2012rg,Petrov:2013nia,Carpenter:2013xra,Berlin:2014cfa}, and electroweakino searches in multi-lepton final states might also be useful in split SUSY scenarios \cite{Gori:2014oua,diCortona:2014yua,diCortona:2015qga}.} Searches for other states that must have particular masses to give the correct relic density may also be important, for example in many coannihilation scenarios the NLSP must be almost degenerate with the dark matter candidate. Alternatively, if a particular SUSY breaking and mediation mechanism is assumed, the ratio of soft masses is fixed, and therefore the 
viable dark matter scenarios can be constrained by the stringent bounds on coloured particles. For common patterns of soft terms, this often gives the dominant constraints \cite{Gori:2014oua,diCortona:2014yua,diCortona:2015qga}.

 The LHC at 8 TeV  is particularly sensitive to scenarios with a Majorana bino coannihilating with light squarks or gluinos in the monojet channel \cite{Aad:2014nra,Aad:2015zva}. Consequently, in the MSSM, bino dark matter is already excluded up to a mass of $350~\GeV$ for squark coannihilation and $770~\GeV$ for gluino coannihilation \cite{deSimone:2014pda}. A 100 TeV collider monojet search with $3000\,\,\mathrm{fb}^{-1}$ could probe bino dark matter almost degenerate with a gluino for masses up to $6.2\,\TeV$, and $4\,\TeV$ for squark coannihilation \cite{Low:2014cba}, significantly constraining gluino-bino and squark-bino coannihilating scenarios. 
On the other hand the bino-slepton scenario is weakly constrained due to the loss of sensitivity when the splitting between the slepton and the bino is small, which is needed to obtain the correct relic density. Future hadron colliders would have difficulty probing the dark matter parameter space, while future lepton colliders would have good sensitivity up to the kinematic limit for almost any mass difference.

In the case of wino or Higgsino LSP the mass splitting between the neutral and the charged states is generated at loop-level. As a result, long-lived charginos can lead to charged tracks in a detector that end when the chargino decays to the LSP \cite{Low:2014cba,Cirelli:2014dsa}. Wino-like neutralinos lighter than $220~\GeV$ are already disfavoured by the LHC long-lived charged wino analysis \cite{Aad:2013yna}, which could reach $500~\GeV$ at a 13 TeV run. The same search at a future 100 TeV hadron collider could constrain wino dark matter masses of $3\,\TeV$, almost saturating the parameter space for pure wino dark matter.

Other searches at the 13 TeV LHC (assuming an integrated luminosity of $3000\,\,\mathrm{fb}^{-1}$) can give significant constraints. The monojet channel could probe 
pure Higgsinos up to $200~\GeV$. Mixed bino-Higgsino and bino-wino neutralino searches through multilepton channel may be probed up to $200~\GeV$. 
On the other hand, a 100 TeV collider could reach Higgsinos with mass slightly less than $1\,\TeV$ using the monojet channel. In addition, mixed bino-Higgsino or bino-wino states, with a mass splitting of $20~\GeV$, could be probed up to masses of $1.3\,\TeV$ through multilepton searches, covering a significant part of the dark matter parameter space in these scenarios.\footnote{These cases are interesting because mixed spectra are able to saturate the relic density for a range of masses, as in the well- tempered neutralino scenario \cite{ArkaniHamed:2006mb} or in the focus point region \cite{Chan:1997bi,Feng:1999zg,Feng:2000gh,Feng:2011aa,Akula:2011jx}.}

 Imposing commonly assumed gaugino mass ratios leads to strong constraints on viable dark matter scenarios. If the gaugino masses are in the unification pattern $M_1:M_2:M_3=\alpha_1:\alpha_2:\alpha_3$, the dark matter candidates are the bino, the Higgsino, or a mixed state of them. LHC bounds on gluinos already rule out bino-Higgsino dark matter up to $210\,\GeV$. A 100 TeV collider may probe gluinos up to $15$ TeV, corresponding to bino masses up to $2.5\,\TeV$. The correct relic density cannot be obtained for such a heavy bino, therefore the only remaining dark matter candidate would be a Higgsino of $1.1\,\TeV$. In models with anomaly mediation, gluino searches at the LHC exclude bino-wino dark matter up to $700 \,\GeV$ with the possibility of reaching up to $3\,\TeV$ for pure wino models at a future 100 TeV collider, similar to the reach of direct searches for long lived charged winos.

\subsection{MRSSM dark matter at colliders}

A Dirac bino LSP coannihilating with first and second generation sleptons is a viable dark matter candidate for masses up to about $300\,\GeV$. The correct relic density is possible for larger mass splittings between a Dirac bino and sleptons than in the Majorana case, and as a result searches for slepton pair production can probe this scenario unlike in the MSSM. 
Because sleptons are directly produced in Drell Yann processes, the collider limits in the slepton--neutralino parameter space obtained by ATLAS and CMS analyses apply to Dirac binos \cite{Aad:2014vma,Khachatryan:2014qwa}, and the ATLAS bound is plotted in Fig.~\ref{fig:BW} right.  
Slepton NLSPs are directly pair produced and subsequently decay into two same flavour leptons and two LSPs (giving rise to missing energy). 
As a result, in this scenario the ATLAS and CMS experiments exclude a bino dark matter lighter than $120\,\GeV$, leaving unexplored regions of small neutralino-slepton mass splitting.
New colliders will be able to improve the reach for sleptons with light LSP mass, but the sensitivity in the interesting region, where $m_{\tilde{l}}-m_{\chi^0_1} < m_W$, is limited by the large background.

The Dirac bino-wino dark matter scenario is complicated to search for at hadron colliders, because the small mass splitting, comparable with the one of the Majorana bino-wino dark matter case, creates issues in background rejection and triggering.
However interesting opportunities arise in boosted electroweakino searches \cite{Bramante:2014tba}. In particular, searches for $pp \to \tilde{\chi}_1^\pm \tilde{\chi}_2^0 \to l^\pm\,\tilde{\chi}_1^0 \gamma\,\tilde{\chi}_1^0\,j$ are very effective in probing models with bino-wino coannihilation. This search has the advantage that the smaller the splitting between the two neutralinos, the greater the branching fraction to photons compared to off-shell $Z$ bosons. The cross section for this process is suppressed at the LHC, but studies for the MSSM at a 100 TeV collider indicate that this search could probe dark matter masses up to $2\,\TeV$ for a luminosity of  $20~ {\rm ab}^{-1}$. The production cross section for charged and neutral Dirac winos is a few times the one in the MSSM, while the splitting between the bino LSP and the wino NLSP is comparable to in the MSSM. Therefore boosted electroweakino studies at a future hadron collider are expected to probe the whole Dirac bino-wino 
dark matter parameter space after a few ab$^{-1}$ of luminosity.

Similarly to the MSSM, if the gaugino mass ratios are fixed, searches for coloured particles can be relevant. If squarks are light compared to a Dirac gluino, LHC searches constrain the gluino mass to be heavier than $1.5\,\TeV$. Therefore, since a Dirac bino must have mass less than about $300\,\GeV$ to be a suitable dark matter candidate, this is only a viable scenario in models with a ratio of gaugino masses $M_{D_3}/M_{D_1} \gtrsim 5$. As seen in Fig.~\ref{fig:Keffect}, for light squarks a 100 TeV collider can probe up to $20\,\TeV$ Dirac gluinos. Consequently, models with Dirac bino dark matter and $M_{D_3}/M_{D_1} \lesssim 70$ will either be discovered or excluded. Given the typical ratios of gaugino masses from UV models (discussed in Section~\ref{sec:model}), for large classes of theories this probes the entire bino dark matter parameter space, surpassing the sensitivity of searches for sleptons.
In the bino-wino scenario, $M_{D_2}$ and $M_{D_1}$ must be close to degenerate. This is not naively the case in any of the simple UV completions considered and therefore the impact of searches for gluinos is unclear.

Finally, searches for squarks and gluinos at 100 TeV colliders can have an interesting interplay with direct detection. In models where the Higgsinos are decoupled, the spin independent cross section is mediated only by squarks. Therefore, LZ can effectively exclude squark masses up to $15\,\TeV$ in models with a Dirac bino LSP. On the other hand, a 100 TeV collider can be sensitive to comparable mass squarks. The discovery of squarks would therefore lead to an expected minimum direct detection cross section in models with bino dark matter, close to the experimentally accessible values. Meanwhile the exclusion of squarks with similar masses would mean that Dirac bino dark matter would have to scatter via a Higgsino component if it was to be visible at LZ.

\section{Conclusions} \label{sec:con}

In this paper we have compared the prospects of discovering or excluding models with Dirac and Majorana gauginos at a possible future 100 TeV collider. Considering a simple model with a neutralino, the first and second generation squark and a gluino, we scanned over squark and gluino masses
to find the potential discovery significance and exclusion reach.  Our results are summarised by Fig.~\ref{fig:majoranadiscovery}. 
Of particular interest are the differences between  Dirac and Majorana gluino models. These are most pronounced for gluinos heavy compared to the squarks, in which case the sensitivity to squarks is significantly weaker in Dirac models. A major shortcoming of the present work is our lack of NLO K factors for Dirac models, and to make accurate predictions this is a 
topic that needs further work. The discovery potential also changes significantly depending on the uncertainty on the signal. Finally, we have not included extra jets at the parton level. Comparing our results to the Majorana study \cite{Cohen:2014hxa} we find this has a non-zero but not enormous effect, and is especially important for large gluino masses.

In extending our analysis there are many other possible searches for colour superpartners, and different patterns of soft masses to be studied. Additionally, study of the scenario where the first two generation sfermions are heavy, but the gluino and stop are relatively light would be worthwhile.  Dedicated analysis of the possible signatures of the sgluon, or the other adjoint scalars, would also be interesting to pursue. It would also be very interesting to consider searches for Dirac winos and binos with or without sfermions at comparable masses and with the gluinos decoupled.
 This is especially relevant since models of Dirac gauginos often predict gluinos significantly heavier than, and winos of roughly similar mass to, the squarks. Collider signatures of electroweakinos in Dirac models have been discussed in \cite{Choi:2008pi,Choi:2010gc}, and the reach of a 100 TeV collider for Majorana gauginos with associated production has been studied \cite{Ellis:2015xba}. However, we leave the case of models of Dirac electroweakinos at a 100 TeV collider to future work.

We have also considered some aspects of model building, and the relation of these to the future collider searches. Although a proper calculation of the fine tuning of a model requires a full UV complete theory to be specified, we have estimated the typical tunings that will be probed in the Dirac and Majorana scenarios. Our analysis has been fairly independent of the details of particular models, and is expected to give a lower bound on the tuning of a theory with a given low scale spectrum. It would therefore be interesting to take a well motivated UV model, for example a theory of Dirac gluinos that avoids tachyonic states, and investigate whether its true tuning is close to our estimates. Another important consideration for model building is the effect of flavour and CP violation from the superpartner spectrum. We have found that while in some parts of parameter space Dirac gauginos alleviate these constraints, they are still severe. Therefore, any model which is to be discovered at a 100 TeV collider 
must still have very particular properties in this sector. This is a significant constraint, especially for models of gravity mediation in which the leading order expectation is $\mathcal{O}(1)$ violation.

Finally, we have examined the prospects for viable neutralino dark matter candidates in Majorana and Dirac models. While it is not the case that a supersymmetric model must have a suitable dark matter candidate, since the dark matter could consist of for example an invisible axion, it is an attractive possibility. 
In R-symmetric Dirac gaugino models, strong constraints from direct detection experiments already rule out Higgsino dark matter. The remaining electroweakino dark matter possibilities are binos coannihilating with sleptons, and a mainly bino neutralino coannihilating with the winos.

LHC searches already constrain the viable bino candidates to be close in mass to the sleptons, and it will be hard for a future hadron collider to strengthen limits in this part of parameter space. However, for many well motivated patterns of soft terms, searches for coloured particles at such a collider can indirectly exclude this scenario. It would be interesting to study whether a future lepton collider could constrain this case directly.
In contrast, the bino-wino case can be fully covered by direct searches. 
Notably, models with Dirac gluinos but a neutralino sector with R-symmetry broken can lead to many other dark matter scenarios. It would be worthwhile to consider the impact of a 100 TeV collider on such theories, as well as R-symmetric models with gravitino dark matter.

\section*{Acknowledgements}
We would like to thank Mike Hance and Kiel Howe for discussion about the Snowmass Standard Model backgrounds and detector simulation. We are also grateful to James Unwin and Giovanni Villadoro for useful discussions.

\appendix

\section{The MRSSM neutralinos}
\label{app:MRSSM}
In this Appendix we briefly review the MRSSM. This has a superpotential  
\bea
W&=& W_{MSSM}^{\mu=0,A=0}+ \mu_d \hat{R}_d \hat{H}_d + \mu_u \hat{R}_u \hat{H}_u \nonumber\\
&+&\Lambda_d \hat{R}_d \hat{T} \hat{H}_d + \Lambda_u \hat{R}_u \hat{T} \hat{H}_u + \lambda_d \hat{S} \hat{R}_d \hat{H}_d + \lambda_u \hat{S} \hat{R}_u  \hat{H}_u ~,
\label{eq:supMRSSM}
\eea
where $\hat{H}_u$ and $\hat{H}_d$ are MSSM-like Higgs doublets, and $\hat{R}_u$ and $\hat{R}_d$ are new inert Higgs doublets added to allow  viable phenomenology while preserving an R-symmetry. The other new fields compared to the MSSM are a singlet $\hat{S}$, a \suw--triplet $\hat{T}$, and a \suc-- octet $\hat{O}$. 

The gauginos $\tilde{B}$, $\tilde{W}$ and $\tilde{g}$  can get, R-symmetry preserving, Dirac mass terms with the fermionic parts of the adjoint chiral supermultiplets $\tilde{S}$, $\tilde{T}$ and $\tilde{O}$
\beq
\mathcal{L}_D \supset M_{D_1}  \tilde{B} \tilde{S} + M_{D_2} \tilde{W} \tilde{T} + M_{D_3} \tilde{g} \tilde{O} + \mathrm{h.c.}~.
\eeq
There are also new trilinear terms involving these extra adjoint chiral multiplets, with coupling constants $\Lambda_{u,d}$ and $\lambda_{u,d}$. Other important parameters are the soft masses for the scalar components of $\hat{H}_{u,d}$ and $\hat{R}_{u,d}$. 
 MSSM like trilinear A  terms are forbidden by the R-symmetry.

During EW symmetry breaking the two MSSM-like Higgs doublets get vacuum expectation values, and $\hat{R}_u$ and $\hat{R}_d$ do not. There are several important differences between the EW sectors of the MSSM and the MRSSM. In the MRSSM, the neutralino mass matrix in the basis $\left(\lambda_{\tilde{B}}, \tilde{W}^0, R_d^0, R_u^0\right), \left(\tilde{S}, \tilde{T}^0, \tilde{H}_d^0, \tilde{H}_u^0\right)$ is

\begin{equation} 
m_{\tilde{\chi}^0} = \left( 
\begin{array}{cccc}
M_{D_1} &0 &-\frac{1}{2} g_1 v_d  &\frac{1}{2} g_1 v_u \\ 
0 &M_{D_2} &\frac{1}{2} g_2 v_d  &-\frac{1}{2} g_2 v_u \\ 
- \frac{1}{\sqrt{2}} \lambda_d v_d  &-\frac{1}{2} \Lambda_d v_d  &m_{R_d^0\tilde{H}_d^0} &0\\ 
\frac{1}{\sqrt{2}} \lambda_u v_u  &-\frac{1}{2} \Lambda_u v_u  &0 &m_{R_u^0\tilde{H}_u^0}\end{array} 
\right) ~,
\label{eq:Mneut}
 \end{equation} 
 where
\begin{align} 
m_{R_d^0\tilde{H}_d^0} &= -\frac{1}{2} \Lambda_d v_T  - \frac{1}{\sqrt{2}} \lambda_d v_s  - \mu_d \\ 
m_{R_u^0\tilde{H}_u^0} &= -\frac{1}{2} \Lambda_u v_T  + \frac{1}{\sqrt{2}} \lambda_u v_s  + \mu_u ~.
\end{align} 
The Dirac neutralinos are therefore composed of eight Weyl spinors coming from the neutral components of the Higgs doublets $H_{u,d}^0$, $R_{u,d}^0$, the gauginos $\tilde{B},\tilde{W}^0$, and the adjoint singlet $\tilde{S}$ and triplet $\tilde{T}$. The four mass parameters $M_{D_1}$, $M_{D_2}$ and $\mu_{u,d}$ give most of the mass to the neutralinos. There are four new couplings $\lambda_{u,d}$ and $\Lambda_{u,d}$, that arise from the superpotential terms in the second line of Eq.~\eqref{eq:supMRSSM}.
Instead in the MSSM the gauginos have Majorana masses and there is only one Higgsino parameter that give mass to two almost degenerate neutralinos.

The neutralino mass matrix is diagonalised by two unitary matrices $N_{ij}^{(k)}$, where $k=1,2$, $i,j=1,...4$.
Conservation of electromagnetic charge and R-charge divides the eight two-component fermions from the winos, Higgsinos and R-fields into four sets that do not mix. This means that there are two chargino mass matrices. Acting on the basis $(\tilde{T}^-,\tilde{H}_d^-),\,\,(\tilde{W}^+,\tilde{R}^+_d)$ we have a mass matrix
\begin{equation} 
m_{\tilde{\chi}^+} = \left( 
\begin{array}{cc}
g_2 v_T  + M_{D_2} &\frac{1}{\sqrt{2}} \Lambda_d v_d \\  
\frac{1}{\sqrt{2}} g_2 v_d  &-\frac{1}{2} \Lambda_d v_T  + \frac{1}{\sqrt{2}} \lambda_d v_s  + \mu_d\end{array} 
\right), 
 \end{equation} 
while on the basis \( \left(\tilde{W}^-, R_u^-\right), \left(\tilde{T}^+, \tilde{H}_u^+\right) \) the mass matrix is
\begin{equation} 
m_{\tilde{\rho}^-} = \left( 
\begin{array}{cc}
- g_2 v_T  + M_{D_2} &\frac{1}{\sqrt{2}} g_2 v_u \\ 
- \frac{1}{\sqrt{2}} \Lambda_u v_u  &-\frac{1}{2} \Lambda_u v_T  - \frac{1}{\sqrt{2}} \lambda_u v_s  - \mu_u \end{array} 
\right)~. 
 \end{equation} 
Both of these matrices is diagonalised by two unitary matrices, so there are four independent rotations.
The parameters $M_{D_1}$, $M_{D_2}$, $\mu_u$ and $\mu_d$ can chosen to be real and positive, because it is possible to rotate any phases into the scalar adjoint holomorphic masses and the parameters $\lambda_{u,d},\,\,\Lambda_{u,d}$.

\bibliography{dirac}
\bibliographystyle{JHEP}

\end{document}